\documentclass[manuscript]{aastex}
\usepackage{epstopdf,natbib}
\usepackage{fullpage}
\usepackage{epstopdf,natbib}
\usepackage{graphicx} 
\usepackage{wrapfig}
\usepackage{amsmath}

\shortauthors{Nugent et al.}
\begin{document}

\title{NEOWISE Reactivation Mission Year One: Preliminary Asteroid Diameters and Albedos}

\author{C. R. Nugent\altaffilmark{1}, A. Mainzer\altaffilmark{2}, J. Masiero\altaffilmark{2}, J. Bauer\altaffilmark{2}, R. M. Cutri\altaffilmark{1}, T. Grav\altaffilmark{3}, E. Kramer\altaffilmark{2}, S. Sonnett\altaffilmark{2}, R. Stevenson, and E. L. Wright\altaffilmark{4}}

\altaffiltext{1}{Infrared Processing and Analysis Center, California Institute of Technology, Pasadena, CA 
91125, USA}
\altaffiltext{2}{Jet Propulsion Laboratory, California Institute of Technology, Pasadena, CA 91109 USA}
\altaffiltext{3}{Planetary Science Institute, Tucson, AZ}
\altaffiltext{4}{Department of Physics and Astronomy, University of California, Los Angeles, CA 90095, USA}

\begin{abstract} 
We present preliminary diameters and albedos for 7,959 asteroids detected in the first year of the NEOWISE Reactivation mission. 201 are near-Earth asteroids (NEAs). 7,758 are Main Belt or Mars-crossing asteroids. 17$\%$ of these objects have not been previously characterized using WISE or NEOWISE thermal measurements.  Diameters are determined to an accuracy of $\sim20\%$ or better. If good-quality H magnitudes are available, albedos can be determined to within $\sim40\%$ or better.
\end{abstract}

\section{Introduction}

Sizes and albedos of asteroids are basic quantities that can be used to answer a range of scientific questions. A significant number of diameter measurements produce a size-frequency distribution, which can constrain models of asteroid formation and evolution \citep{1979aste.book..783Z,1982Sci...216.1405G,2002AJ....123.1056T,2002Icar..158..146B,Masiero11}. Asteroid albedos aid the identification of collisional family members \citep{Masiero.2013a, Walsh.2013a, Carruba.2013a, Milani.2014a,MasieroAstIV_temp}, and allow for basic characterization of asteroid composition \citep{Mainzer2011b, Grav12, Masiero14}.

Most observations of asteroids are made in visible wavelengths, where flux is dependent on both size and albedo. Observations in other wavelengths, such as the infrared \citep[e.g.][]{Hansen.1976a, Cruikshank.1977a, Lebofsky.1978a, Morrison.1979a, Delbo.2003a, Delbo.2011a, Matter11, Mueller.2012a, Mueller.2013a, Wolters.2005a, Wolters.2008a} or radio \citep[e.g.][]{Ostro.2002a,BennerAstIV_temp}, are needed to determine these quantities precisely.  At present, well-determined diameters and albedos have been measured for less than a quarter of known asteroids. 

The infrared NEOWISE project \citep{Mainzer2011a} has measured diameters and albedos for $\sim20\%$ of the known asteroid population, the majority of these measurements to date \citep{Mainzer2011d,Mainzer12,MainzerAstIV_temp,Masiero11,Masiero12, GravTrojans, GravHilda, GerbsCentaur}. Here, we expand the number of asteroids characterized by NEOWISE, deriving diameters and albedos for asteroids detected by NEOWISE between December 13, 2013, to December 13, 2014 during the first year of the Reactivation mission.

The NEOWISE mission uses the Wide-Field Infrared Survey Explorer (WISE) spacecraft, which images the entire sky using freeze-frame scanning from a sun-synchronous polar orbit \citep{Wright10WISE, Cutri12}.
WISE is equipped with a
50 cm telescope and four 1024x1024 pixel focal plane array detectors
that simultaneously image the same 47x47 arc minute field-of-view
in 3.4, 4.6, 12 and 22 $\mu$m bands,  all originally cooled by solid hydrogen
cryogen.  WISE scans the sky between the ecliptic poles continuously
during its 95 minute orbit. A tertiary scan mirror freezes the sky on the 
focal planes for 11 seconds while the detectors are read out, producing
a sequence of adjacent images with 7.7 sec exposure times in 3.4 and 
4.6 $\mu$m bands and 8.8 sec in 12 and 22 $\mu$m bands.  The orbit precesses at 
an average rate of approximately one degree per day, so that the full
sky is covered in six months.

WISE was launched on December 14, 2009, and began surveying  on January 7, 2010.  WISE scanned the sky 1.5 times during the 9.5 months while it was cooled by its hydrogen cryogen. After the hydrogen was
depleted, the survey continued as NEOWISE until February 1, 2011 using 
the 3.4 and 4.6 $\mu$m detectors that operated at near full sensitivity
with purely passive cooling.  During the additional four months of
``post-cryo'' operations, coverage of the entire inner Main Asteroid Belt was completed, along with a second complete coverage of the sky.  WISE/NEOWISE was placed into hibernation in mid-February 2011.  In this mode,
the solar panels were held facing the Sun and the telescope pointed 
towards the north ecliptic pole.  The telescope viewed the Earth
during half of each orbit, resulting in some heating.

The WISE spacecraft was brought out of hibernation in September 2013
and renamed NEOWISE to continue its mission to discover, track and 
characterize asteroids through $\sim2017$ \citep{14MainzerRestart}.  The telescope was restored to near zenith pointing,
which enabled the optics and focal planes to cool passively back to $\sim74$K.
Survey operations resumed on December 13, 2013, with the 3.4 and 4.6
$\mu$m detectors operating at a sensitivity comparable to that during the
original WISE cryogen survey \citep{2015Cutri}. The NEOWISE moniker, the acronym of Near-Earth Object WISE, encompasses both the archiving of individual images, to allow for the detection of transient objects, and the extensions of the mission beyond WISE's original 9-month 
lifetime.

NEOWISE uses the same survey and observing strategy as the
original WISE mission \citep{Wright10WISE}.  The majority of each orbit is devoted to observations, with only brief breaks for data transmission and momentum unloading. The spacecraft carries a body-fixed antenna, and therefore must reorient itself to communicate with the Tracking and Data Relay Satellite System (TDRSS), which relays the data to Earth. 
Data transmission is timed to only interrupt survey coverage near the ecliptic poles, which are observed frequently. Momentum unloading, which can result in streaked images, is also completed at this time.

Data processing for NEOWISE uses the WISE
Science Data System (Cutri et al. 2015) that performs instrumental,
photometric and astrometric calibration for each individual set of 3.4 and 4.6
$\mu$m exposures obtained by the spacecraft, and detects and characterizes 
sources on each exposure.  The calibrated images and the database of 
positions and fluxes of sources extracted from those images for the
first year of NEOWISE survey observations were released in March 2015 \citep{2015Cutri}.

The WISE Moving Object Pipeline System \citep[WMOPS;][]{Cutri12} identifies sources that display motion between 
the different observations of the same region on the sky.  WMOPS uses 
the extracted source lists from sets of images to first identify and filter out sources that appear stationary between individual exposures, and then links non-stationary detections into sets that exhibit physically plausible motion on
the sky. Generally, objects within 70 AU of the sun move quickly enough to be detected by WMOPS (\citealt{Mainzer2011a}, see also \citealt{GerbsCentaur}). Those candidate moving objects that are not associated with known asteroids, comets, planets or planetary satellites are verified individually by NEOWISE scientists.  A minimum of five independent detections are required for a ÒtrackletÓ (a set of position/time pairs) to
be considered reliable.  Tracklets for each verified new candidate object and previously known solar system objects are reported to the IAU Minor Planet Center (MPC) three times per week.  The MPC performs initial orbit determination, associates the NEOWISE tracklets with known objects, and archives the NEOWISE astrometry and times in its observation database.

Candidates confirmed by the MPC to be possible new near-Earth objects (NEOs) are posted to their NEO Confirmation page for prompt follow-up observations
by ground-based observers.  Rapid follow-up is essential for NEOWISE
NEO candidates because the NEOWISE arcs are usually short, and the asteroid's projected positional uncertainties grow quickly, making reliable
recovery difficult after 2-3 weeks. To ensure prompt follow-up, NEOWISE observations are reported to the MPC less than three days after observations on board the spacecraft. A NEOWISE candidate discovery has a minimum of $5$ observations over $\sim3$ hours, although typical objects have $\sim12$ observations spanning $\sim1.5$ days.

Targets observed by NEOWISE can pose unique challenges to ground based follow-up observers. NEOWISE's orbit allows observations to be made at all declinations, and observing is independent of lunar phase. Ground-based observers are limited to a fixed declination range, and must sometimes deal with light from the moon and terrestrial weather, which can preclude observations. Moreover, NEOWISE discoveries are frequently extremely dark (see Figure \ref{fig:disc}), often requiring 2-4 m class telescopes to detect them at low solar elongations.  

\begin{figure}[h!]
  \caption{Number of follow-up observations by observatories that contributed $>5$ observations during the Year 1 Reactivation. Spacewatch, LCOGT, and Catalina employ multiple telescopes; their observatory codes have been grouped together. Observatory code 568, Mauna Kea, is frequently used by the Tholen group.}
  \centering
  \includegraphics[width=1.0\textwidth]{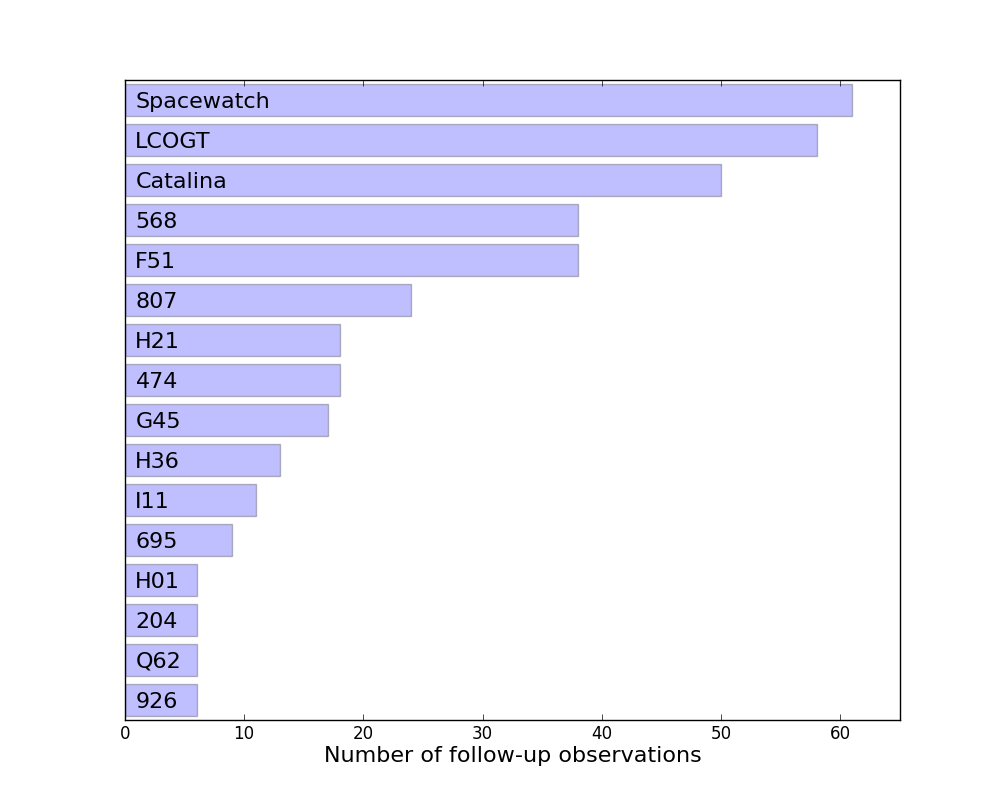}
  \label{fig:followup}
\end{figure}

Observers around the globe (including both amateurs and professionals) have contributed essential follow-up observations, which are defined here as an observation of an object within $15$ days of its first observation on board the spacecraft. Significant contributors of follow-up observations are given in Figure \ref{fig:followup}. The Spacewatch Project \citep{2007McMillan} contributes a large share of recoveries in the northern hemisphere. The Las Cumbres Observatory Global Telescope (LCOGT) Network of robotically operated queue-scheduled telescopes \citep{Brown2013} has been an extremely useful resource for securing detections when weather is poor at a particular site.   Additionally, the group led by D. Tholen using the University of Hawaii 2.2 m and Canada-France-Hawaii 4 m telescopes has successfully detected the targets with the faintest optical magnitudes in the northern hemisphere \citep[e.g.][]{2014MPECTholen}.  The NEOWISE team was awarded time with the DECam instrument on the Cerro Tololo Inter-American Observatory (CTIO) 4m telescope, which has proven invaluable for the recovery of low albedo objects at extreme declinations in the southern hemisphere.

We present diameters and albedos for 201 near-Earth asteroids (NEAs) and 7,758 Main Belt and Mars-crossing asteroids detected in the first year of reactivation, between December 13, 2013, and December 13, 2014. This includes the 38 NEAs discovered by NEOWISE during those dates.

\section{Methods}
\subsection{Observations}

The MPC is responsible for verifying and archiving asteroid astrometry. To obtain the verified record of objects found by the WMOPS pipeline in the NEOWISE data, we queried the MPC observations files `NumObs.txt' and `UnnObs.txt' for NEOWISE (observatory code C51) observations between December 13, 2013 and December 13, 2014. This returned the list of object
identifications, along with the observation times and NEOWISE measured astrometry. This included known objects and WMOPS asteroid discoveries made during that time.

The NASA/IPAC Infrared Science Archive (IRSA, at http://irsa.ipac.caltech.edu) NEOWISE-R Single Exposure Source Table was then queried for the fluxes of sources detected in the NEOWISE data. The list of detections extracted from the MPC files was converted into GATOR format (see http://irsa.ipac.caltech.edu/applications/Gator/GatorAid/irsa/QuickGuidetoGator.htm), and uploaded into the IRSA interface using a cone search radius of 2 arcsec and a restriction that times match the MPC-archived observation time to within 2 seconds. This two-step process of querying both the MPC archive and the NEOWISE-R Single Exposure (L1b) Source Table ensures that only detections verified both by the NEOWISE object identification routines and the MPC are used for thermal modeling. While there may be additional objects in the database that were detected fewer than 5 times, or are just below the single-frame detection threshold, this method of extracting moving object detections ensures high reliability, since WMOPS actively works to exclude fixed sources such as stars and galaxies from tracklets.  Sources with fewer than 5 detections or those that fall just below the single-frame detection threshold will be extracted in future processing.

NEOWISE detections were further filtered using several measurement and image quality flags. We required detections to have ``ph\_qual'' values of  ``A'', ``B'', or ``C'', ``cc\_flag'' values of ``0'', and ``qual\_frame'' values of ``10''. The `ph\_qual'' flag represents photometric quality, accepting a value of ``C'' or higher ensures that the sources was detected with a flux signal-to-noise ratio $>2$. The ``cc\_flag'', or contamination and confusion flag, indicates whether the source measurement may be compromised due to a nearby image artifact. By filtering for ``cc\_flag''$=0$, we select for sources unaffected by known artifacts. Finally, ``qual\_frame'' is an overall quality grade for the entire image in which the source was detected. We accepted only the best-quality images, those with a score of ``10''.

The filtered data from the NEOWISE Single Exposure Source Table are high-quality source measurements that were found at the times and locations of NEOWISE WMOPS detections submitted to the MPC. To further guard against the possibility of confusing a minor planet with fixed background sources such as stars and galaxies, we uploaded the filtered data to the IRSA catalog query engine, referencing the WISE All-Sky Source Catalog to determine if any single-frame detections fell within 6.5 arcsec of an Atlas source. The WISE Source Catalog is generated using multiple independent single exposure images. Fast-moving solar system objects are suppressed during the construction of the catalog. A search radius of 6.5 arcsec was chosen as it is the approximate size of the WISE beam in the 3.4 and 4.6 $\mu$m bands.

We required at least three observations with magnitude errors $\sigma_{mag} \le 0.25$ in one band. The largest main-belt asteroids (MBAs) can saturate the NEOWISE detectors, resulting in reduced photometric accuracy. Following the prescription laid out in \citet{Cutri12} (Section IV.4), we did not consider objects that were brighter than 8.0 mag at $3.4 \mu$m and 7.0 mag at $4.6 \mu$m.  The NEA measurements used in this work are given in Table \ref{tab:obsshort}.

\begin{deluxetable}{rrrrr}
\tabletypesize{\scriptsize}

\tablecaption{NEOWISE magnitudes for the NEAs modeled in this paper.  Given are the time of the observation in modified Julian date (MJD), and the magnitude in the $3.4\mu$m (W1) and $4.6\mu$m bands (W2). Non-detections at a particular wavelength represent $95\%$ confidence limits \citep{Cutri12}. The aperture radius in arcsec used for
aperture photometry is given under ``Aperture"; ``0" indicates that the pipeline profile fit photometry was used. Observations for the first three objects only are shown; the remainder are available in electronic format through the journal website.}
\tablewidth{0pt}
\tablehead{
	\colhead{Name} & \colhead{MJD} & \colhead{W1 (mag)}& \colhead{W2 (mag)}& \colhead{Aperture} }                                                                                   
\startdata
  01566 & 56795.5373147 & $>$16.339 & 13.317 $\pm$ 0.086 &  0\\  
   01566 & 56795.668982 & 15.340 $\pm$ 0.132 & 13.287 $\pm$ 0.104 &  0\\  
   01566 & 56795.8005219 & 15.270 $\pm$ 0.133 & 13.255 $\pm$ 0.157 &  0\\  
   01566 & 56795.8663555 & 15.268 $\pm$ 0.137 & 13.226 $\pm$ 0.125 &  0\\  
   01566 & 56795.9321892 & 15.590 $\pm$ 0.200 & 13.395 $\pm$ 0.166 &  0\\  
   01566 & 56796.1295626 & 14.904 $\pm$ 0.097 & 13.348 $\pm$ 0.102 &  0\\  
   01566 & 56796.2612299 & 15.829 $\pm$ 0.192 & 13.467 $\pm$ 0.196 &  0\\  
   01580 & 56955.905682 & $>$16.484 & 14.033 $\pm$ 0.171 &  0\\  
   01580 & 56956.037222 & $>$16.124 & 14.230 $\pm$ 0.156 &  0\\  
   01580 & 56956.431715 & 17.100 $\pm$ 0.538 & 13.972 $\pm$ 0.136 &  0\\  
   01580 & 56956.5631277 & 16.951 $\pm$ 0.474 & 14.158 $\pm$ 0.198 &  0\\  
   01580 & 56956.6289614 & $>$16.168 & 14.159 $\pm$ 0.157 &  0\\  
   01580 & 56956.6946677 & 16.178 $\pm$ 0.252 & 14.312 $\pm$ 0.187 &  0\\  
   01580 & 56956.7603741 & 16.442 $\pm$ 0.316 & 13.976 $\pm$ 0.154 &  0\\  
   01580 & 56956.8262078 & $>$17.166 & 13.988 $\pm$ 0.209 &  0\\  
   01580 & 56956.8919142 & 16.944 $\pm$ 0.523 & 14.050 $\pm$ 0.134 &  0\\  
   01580 & 56956.9576205 & 16.206 $\pm$ 0.291 & 14.282 $\pm$ 0.186 &  0\\  
   01580 & 56957.0891606 & 16.795 $\pm$ 0.397 & 14.271 $\pm$ 0.180 &  0\\  
   01580 & 56957.4179471 & $>$17.009 & 13.987 $\pm$ 0.145 &  0\\  
   01620 & 56993.9087248 & 15.427 $\pm$ 0.137 & 14.075 $\pm$ 0.156 &  0\\  
   01620 & 56994.3030911 & 15.463 $\pm$ 0.154 & 14.049 $\pm$ 0.200 &  0\\  
   01620 & 56994.434504 & 15.420 $\pm$ 0.144 & 13.556 $\pm$ 0.102 &  0\\  
   01620 & 56994.5659171 & 15.596 $\pm$ 0.171 & 14.375 $\pm$ 0.205 &  0\\  
   01620 & 56994.5660444 & 16.012 $\pm$ 0.212 & 14.305 $\pm$ 0.221 &  0\\  
   01620 & 56994.6317509 & 15.754 $\pm$ 0.198 & 13.846 $\pm$ 0.169 &  0\\  
   01620 & 56994.7631639 & 15.513 $\pm$ 0.145 & 13.807 $\pm$ 0.132 &  0\\  
   01620 & 56994.8945768 & 15.794 $\pm$ 0.216 & 14.203 $\pm$ 0.228 &  0\\  
   01620 & 56994.8947042 & 15.843 $\pm$ 0.488 & 14.106 $\pm$ 0.155 &  0\\  
   01620 & 56994.9604107 & 15.241 $\pm$ 0.129 & 13.988 $\pm$ 0.155 &  0\\  
   01620 & 56995.0918237 & 15.203 $\pm$ 0.109 & 13.637 $\pm$ 0.140 &  0\\  
   01620 & 56995.2890705 & 15.354 $\pm$ 0.124 & 13.483 $\pm$ 0.115 &  0\\  
   01620 & 56995.4204835 & 15.411 $\pm$ 0.161 & 13.768 $\pm$ 0.175 &  0\\  
   01620 & 56995.8147225 & 15.912 $\pm$ 0.223 & 14.289 $\pm$ 0.205 &  0\\  
\enddata
\label{tab:obsshort}
\end{deluxetable}

\subsection{NEATM}

We used the Near-Earth Thermal Model (NEATM) of \citet{Harris98}, following the implementation of \citet{Mainzer2011d,Mainzer12} for NEAs and \citet{Masiero11,Masiero12} for MBAs and Mars crossers. These results supersede those published in \citet{14MainzerRestart}. NEATM is a simple but effective method for determining effective spherical diameters and albedos (when corresponding visible light observations are available). This model makes several assumptions, including a spherical, non-rotating body, with a simple temperature distribution:

\begin{equation}
T(\theta)=T_{max} \cos^{1/4}(\theta) \quad \textrm{for} \quad 0 \leq \theta \leq \pi/2
\label {eq:NEATM}
\end{equation}

where $\theta$  is the angular distance from the sub-solar point. $T_{max}$ is the subsolar temperature, defined as:
\begin{equation}
T_{max}=\left( \frac{(1-A)S}{\eta \epsilon \sigma}\right)^{1/4}
\end{equation}
where $A$ is the bolometric Bond albedo, $S$ is the solar flux at the asteroid, $\eta$ is termed the beaming parameter, $\epsilon$ is the emissivity, and $\sigma$ is the Stefan-Boltzmann constant. The beaming parameter $\eta$ accounts for any deviation between the actual asteroid and the model. Changes in $\eta$ can account for a host of factors including non-spherical shapes, the presence of satellites, variations in surface roughness or thermal inertia, uncertainties in emissivity, high rates of spin, changes in surface temperature distributions due to spin pole location, or the imprecise assumption that the object's night-side has zero thermal emission (a factor that is most relevant for objects observed at high phase angles). Some of these factors accounted for in the beaming parameter are degenerate. For example, a slow-rotating object will have a heat distribution similar to a faster rotating object that has a lower thermal inertia.  For this simple model, beaming accounts for the changes in temperature distribution due to these effects that cannot be otherwise separated.

Observations were divided into apparitions of 10 days, and the NEATM model was fitted to each individual apparition. These shorter apparitions allowed for fits to widely-spaced apparitions or, for NEAs, over changing phase angles. Given that the NEOWISE observational cadence generally results in an object being detected over $\sim1.5$ days, sometimes with an additional epoch of observations $\sim3-6$ months later, we chose to divide observations separated by $>10$ days for separate fitting to account for large changes in object distances and viewing geometries.

NEATM spheres were approximated by a faceted polygon with 800 facets. Individual facet temperature was determined following Equation \ref{eq:NEATM}, and then color corrected following \citet{Wright10WISE}. Observed thermal flux for each facet was computed, as was flux from reflected sunlight. The integrated flux from the object was determined, accounting for viewing geometry, to produce a model magnitude. A least-squares fitting routine compared modeled to observed magnitudes, and iterated on diameter, albedo, and beaming parameter until a best fit was found.  

Geometric optical albedo $p_V$ was computed using absolute magnitude $H$ and slope parameter $G$, using values supplied in MPCORB.dat by the MPC. Inaccurate $H$ and $G$ values will result in inaccurate $p_V$ fits.  Work by \citet{Williams.2012a} and \citet{Pravec.2012a} found systematic $H$ offsets that vary as a function of H magnitude in data reported to the MPC. As albedo measurements depend on $H$ and $G$ values, errors in measurement of those values will propagate to derived albedos. 

\begin{figure}[h!]
  \caption{Comparison of spectral energy distribution for a simulated NEO and inner main-belt asteroid, each with albedos ranging from $p_{V}=0.06$ to $p_V=0.5$. Thick lines show the flux from the asteroid as a function of wavelength, which is composed of both thermally emitted (dashed) and reflected sunlight (dotted) components. NEOWISE bands centered at 3.4 and 4.6 $\mu$m are shown as shaded cyan and purple bars, respectively. For NEAs (left), the 3.4 and 4.6 $\mu$m bands are both thermally dominated. For objects in the inner Main Belt (right), the 3.4 $\mu$m band is dominated by reflected light, and the 4.6 $\mu$m band is dominated by thermally emitted light, though the ratio between these components varies with albedo.}
  \centering
  \includegraphics[width=1.0\textwidth]{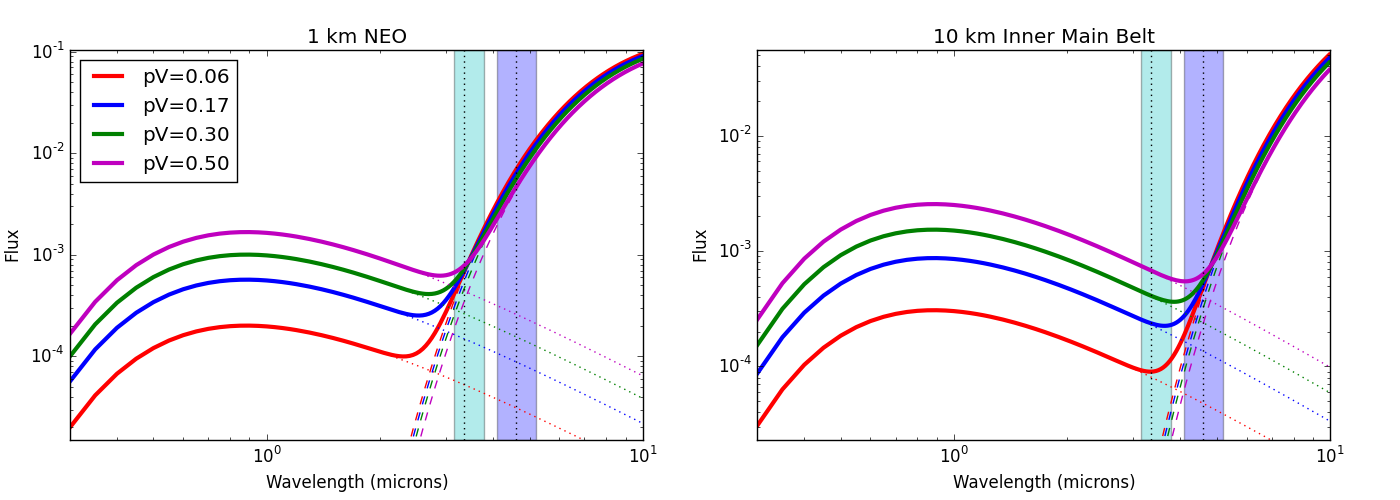}
  \label{fig:sed}
\end{figure}

NEATM requires at least one of the NEOWISE wavelengths to be dominated by thermally emitted light. Some outer main-belt objects observed by NEOWISE were too cold to have thermally dominated emission at 3.4 or 4.6 $\mu$m, and therefore diameters and albedos for those objects are not reported in this paper. The proportion of reflected vs. thermally emitted light for NEAs and inner MBAs can be seen in the spectral energy distribution plots shown in Figure \ref{fig:sed}. The proportion of thermally emitted flux depends on albedo, which means that for colder, outer MBAs it is unclear if a wavelength is thermally dominated until after the fit was performed. Comparison of those results to NEOWISE fits using 12 micron images and radar data confirmed that the thermal fits were poor, so we did not include results that had more than $25\%$ reflected light in the $4.6\mu$m band.

We assumed that $\eta$ was equal to the average value for the object's population, as determined by \citet{Mainzer2011d} or \citet{Masiero11}, respectively. For NEAs, this meant $\eta=1.4\pm0.5$; for all other asteroids in this paper, $\eta=0.95\pm0.25$. As shown in \citet{Masiero11}, although the average $\eta$ for the main belt is 1.0, the peak of the histogram is located closer to $0.95$, so this value was adopted in this work.

Following the average values determined by \citet{Mainzer2011d} and \citet{Masiero11}, the ratio of infrared to visible albedo $p_{IR}/p_{V}$ was initially set to $1.6 \pm 1.0$ for NEAs and $1.5 \pm 0.5$ for Mars-crossers and MBAs. Additionally, it was assumed that the albedos of each band were equal, or  $p_{3.4 \mu m} = p_{4.6 \mu m}$. Although this may be a poor assumption for objects with red slopes \citep{GravTrojans2}, it is necessary to prevent over fitting of the data.

\subsection{Uncertainties}

Uncertainties on $d$, $p_V$, and $\eta$ (when $\eta$ was a free parameter) were determined using a Monte Carlo method. Measured NEOWISE magnitudes, $H$ and $G$ were randomly adjusted within their errors, and the resultant model values of $d$, $p_V$, and (in appropriate cases) $\eta$, were compared to the best-fit values. This process was repeated 25 times for each object, and the resultant errors are the weighted standard deviation of the Monte Carlo trials. The errors quoted in the tables below only include the random component measured through this MC method, not the systematic offset.

Systematic errors were computed by comparing the match between diameters derived in this work to radar-derived diameters for the same objects. Albedos were derived from the radar diameters using the equation 
\begin{equation}
d=\frac{1329}{\sqrt{p_V}}10^{-H/5}
\end{equation}
where $d$ is the diameter in kilometers \citep[for more information, see][]{HarrisLagerrosIII}.

\subsection{Objects without visible-wavelength detections}

Some MBA and Mars-crossing asteroids had no visible-wavelength measurements available from the MPC. Unlike NEAs, objects determined to have these orbits by the MPC are not added to the MPC's NEO Confirmation Page. Therefore, optical follow-up of these objects is rare, and usually serendipitous. For objects without reported optical detections, the $H$ values in MPCORB.dat represent estimates, not measurements, and $p_V$ could not be derived. Since thermally emitted light weakly depends on albedo, $d$ measurements are reported for these objects. However, lacking targeted follow-up, these objects have short arcs and relatively large position uncertainties, which can add additional systematic errors to the derived diameters. 

\subsection{NEAs}

NEAs were examined with particular care. Objects with bad matches between observed and modeled $H$ values were refitted with a parameter that tightened the constraints between modeled and observed H.  Finally, in some cases an assumption of fixed $\eta=1.4$ produced a poor result. For NEAs with poor fits, beaming was varied between 1.0 and 2.0, to see if a statistically significant improvement in fit to the observed NEOWISE magnitudes could be achieved.

\section{Results}

Results are divided into four tables. As diameters were calculated using different parameters for the NEAs vs the MBAs and Mars-crossing asteroids, results for these two groups are presented separately. Results are further subdivided between objects that were characterized previously by the NEOWISE team, and objects that were not. This is because previously published values likely used the fully cryogenic 12 and 22 $\mu$m wavelengths, and therefore can derive diameters more accurately, to within $10\%$. Researchers looking for the best-constrained diameter and albedo measurements should consult previously published work \citep{Mainzer2011d,Masiero11,Mainzer12,Masiero12}. However, for those researchers who are interested in diameters and albedos derived from additional epochs of data provided by the Year One Reactivation results, we also include the diameters derived for objects using these new data. 

Tables \ref{tab:neonew} and \ref{tab:neodup} contain the fit diameters and albedos for 173 new and 28 previously characterized NEAs, respectively. Tables \ref{tab:mbanew} and \ref{tab:mbadup} contain the fit diameters and albedos for 1,176 new and 6,579 previously characterized MBAs and Mars crossing asteroids, respectively. Several objects were observed at multiple apparitions; in these cases, results are presented for each apparition.

\begin{deluxetable}{rrrrrrrrrr}
\tabletypesize{\scriptsize}

\tablecaption{Measured diameters ($d$) and albedos ($p_V$) of near-Earth objects not previously characterized using NEOWISE data. Magnitude $H$, slope parameter $G$, and beaming $\eta$ used are given. The numbers of observations used in the 3.4 $\mu$m ($n_{W1}$) and 4.6 $\mu$m ($n_{W2}$) wavelengths are also reported, along with the amplitude of the 4.6 $\mu$m light curve (W2 amp.).}
\tablewidth{0pt}
\tablehead{
	\colhead{Name} & \colhead{Packed} & \colhead{$H$}& \colhead{$G$}& \colhead{$d$ (km)} & \colhead{$p_V$} & \colhead{$\eta$} & \colhead{W2 amp.} & \colhead{$n_{W1}$} & \colhead{$n_{W2}$} }                                                                                     
\startdata
 1566           & 01566 & 16.90 & 0.15 & 1.03 $\pm$ 0.04 & 0.29 $\pm$ 0.05 & 1.40 $\pm$ 0.00 & 0.24 & 5 & 7 \\ 
 1580           & 01580 & 14.50 & 0.15 & 8.55 $\pm$ 5.23 & 0.04 $\pm$ 0.08 & 1.40 $\pm$ 0.52 & 0.34 & 0 & 12 \\ 
 1620           & 01620 & 15.60 & 0.15 & 1.87 $\pm$ 0.05 & 0.29 $\pm$ 0.04 & 1.40 $\pm$ 0.00 & 0.89 & 14 & 14 \\ 
 1862           & 01862 & 16.25 & 0.09 & 1.40 $\pm$ 0.04 & 0.29 $\pm$ 0.04 & 1.40 $\pm$ 0.00 & 0.41 & 10 & 10 \\ 
 1862           & 01862 & 16.25 & 0.09 & 1.26 $\pm$ 0.04 & 0.35 $\pm$ 0.05 & 1.40 $\pm$ 0.00 & 0.84 & 10 & 10 \\ 
 1917           & 01917 & 13.90 & 0.15 & 4.99 $\pm$ 0.14 & 0.20 $\pm$ 0.03 & 1.40 $\pm$ 0.00 & 0.58 & 14 & 14 \\ 
 1943           & 01943 & 15.75 & 0.15 & 2.34 $\pm$ 0.05 & 0.16 $\pm$ 0.02 & 1.40 $\pm$ 0.00 & 0.22 & 30 & 31 \\ 
 1943           & 01943 & 15.75 & 0.15 & 2.30 $\pm$ 0.04 & 0.17 $\pm$ 0.02 & 1.40 $\pm$ 0.00 & 0.29 & 171 & 172 \\ 
 1943           & 01943 & 15.75 & 0.15 & 2.28 $\pm$ 0.05 & 0.17 $\pm$ 0.03 & 1.40 $\pm$ 0.00 & 0.54 & 14 & 15 \\ 
 2062           & 02062 & 16.80 & 0.15 & 0.80 $\pm$ 0.03 & 0.52 $\pm$ 0.10 & 1.40 $\pm$ 0.00 & 0.82 & 32 & 36 \\ 
 3288           & 03288 & 15.20 & 0.15 & 2.49 $\pm$ 0.07 & 0.24 $\pm$ 0.04 & 1.40 $\pm$ 0.00 & 1.40 & 11 & 11 \\ 
 4954           & 04954 & 12.60 & 0.15 & 9.56 $\pm$ 0.24 & 0.18 $\pm$ 0.03 & 1.40 $\pm$ 0.00 & 1.06 & 8 & 8 \\ 
 5381           & 05381 & 16.50 & 0.15 & 0.91 $\pm$ 0.05 & 0.54 $\pm$ 0.07 & 1.40 $\pm$ 0.00 & 0.49 & 10 & 10 \\ 
 5381           & 05381 & 16.50 & 0.15 & 0.94 $\pm$ 0.04 & 0.51 $\pm$ 0.06 & 1.40 $\pm$ 0.00 & 0.17 & 13 & 13 \\ 
 6053           & 06053 & 14.90 & 0.15 & 3.72 $\pm$ 0.08 & 0.14 $\pm$ 0.02 & 1.40 $\pm$ 0.00 & 0.21 & 11 & 11 \\ 
 7025           & 07025 & 18.30 & 0.15 & 0.50 $\pm$ 0.17 & 0.34 $\pm$ 0.23 & 1.40 $\pm$ 0.52 & 0.58 & 0 & 4 \\ 
 7889           & 07889 & 15.20 & 0.15 & 1.68 $\pm$ 0.07 & 0.52 $\pm$ 0.06 & 1.40 $\pm$ 0.00 & 0.45 & 8 & 8 \\ 
 8567           & 08567 & 15.30 & 0.15 & 2.93 $\pm$ 0.07 & 0.16 $\pm$ 0.03 & 1.40 $\pm$ 0.00 & 0.42 & 25 & 26 \\ 
 13651           & 13651 & 17.60 & 0.15 & 0.56 $\pm$ 0.02 & 0.51 $\pm$ 0.11 & 1.40 $\pm$ 0.00 & 1.22 & 11 & 11 \\ 
 35107           & 35107 & 16.80 & 0.15 & 0.91 $\pm$ 0.03 & 0.41 $\pm$ 0.05 & 1.40 $\pm$ 0.00 & 0.25 & 10 & 10 \\ 
 35107           & 35107 & 16.80 & 0.15 & 1.10 $\pm$ 0.28 & 0.28 $\pm$ 0.16 & 1.40 $\pm$ 0.37 & 0.45 & 0 & 14 \\ 
 39572           & 39572 & 16.50 & 0.15 & 1.55 $\pm$ 0.66 & 0.18 $\pm$ 0.16 & 1.40 $\pm$ 0.47 & 0.42 & 0 & 8 \\ 
 39796           & 39796 & 15.70 & 0.15 & 2.13 $\pm$ 0.59 & 0.20 $\pm$ 0.20 & 1.40 $\pm$ 0.39 & 0.69 & 0 & 16 \\ 
 53430           & 53430 & 16.60 & 0.15 & 1.23 $\pm$ 0.32 & 0.27 $\pm$ 0.15 & 1.40 $\pm$ 0.37 & 1.14 & 0 & 5 \\ 
 54686           & 54686 & 16.50 & 0.15 & 1.35 $\pm$ 0.46 & 0.24 $\pm$ 0.19 & 1.40 $\pm$ 0.47 & 1.02 & 0 & 10 \\ 
 55532           & 55532 & 16.10 & 0.15 & 1.31 $\pm$ 0.04 & 0.38 $\pm$ 0.06 & 1.40 $\pm$ 0.00 & 0.22 & 6 & 6 \\ 
 68063           & 68063 & 15.50 & 0.15 & 2.30 $\pm$ 0.07 & 0.21 $\pm$ 0.04 & 1.40 $\pm$ 0.00 & 0.38 & 24 & 24 \\ 
 68267           & 68267 & 16.90 & 0.15 & 0.88 $\pm$ 0.04 & 0.40 $\pm$ 0.05 & 1.40 $\pm$ 0.00 & 0.29 & 13 & 15 \\ 
 68348           & 68348 & 14.20 & 0.15 & 3.51 $\pm$ 0.13 & 0.30 $\pm$ 0.05 & 1.40 $\pm$ 0.00 & 0.46 & 12 & 12 \\ 
 68548           & 68548 & 16.50 & 0.15 & 1.18 $\pm$ 0.04 & 0.32 $\pm$ 0.04 & 1.40 $\pm$ 0.00 & 0.24 & 8 & 10 \\ 
 68548           & 68548 & 16.50 & 0.15 & 1.24 $\pm$ 0.04 & 0.29 $\pm$ 0.03 & 1.40 $\pm$ 0.00 & 0.55 & 23 & 24 \\ 
 85182           & 85182 & 17.10 & 0.15 & 1.03 $\pm$ 0.37 & 0.24 $\pm$ 0.19 & 1.40 $\pm$ 0.49 & 0.69 & 0 & 9 \\ 
 85774           & 85774 & 19.20 & 0.15 & 0.94 $\pm$ 0.01 & 0.04 $\pm$ 0.01 & 1.40 $\pm$ 0.00 & 0.90 & 11 & 11 \\ 
 86819           & 86819 & 17.40 & 0.15 & 0.80 $\pm$ 0.27 & 0.30 $\pm$ 0.22 & 1.40 $\pm$ 0.46 & 0.79 & 0 & 7 \\ 
 86829           & 86829 & 15.90 & 0.15 & 1.43 $\pm$ 0.05 & 0.37 $\pm$ 0.05 & 1.40 $\pm$ 0.00 & 0.33 & 14 & 14 \\ 
 87309           & 87309 & 17.60 & 0.15 & 0.57 $\pm$ 0.16 & 0.50 $\pm$ 0.23 & 1.40 $\pm$ 0.47 & 0.67 & 0 & 10 \\ 
 88213           & 88213 & 19.70 & 0.15 & 0.91 $\pm$ 0.42 & 0.03 $\pm$ 0.03 & 1.40 $\pm$ 0.51 & 0.66 & 0 & 6 \\ 
 89355           & 89355 & 15.60 & 0.15 & 2.04 $\pm$ 0.05 & 0.25 $\pm$ 0.03 & 1.40 $\pm$ 0.00 & 1.19 & 30 & 31 \\ 
 90075           & 90075 & 15.20 & 0.15 & 2.23 $\pm$ 0.08 & 0.29 $\pm$ 0.04 & 1.40 $\pm$ 0.00 & 0.73 & 12 & 12 \\ 
 99248           & 99248 & 16.30 & 0.15 & 1.12 $\pm$ 0.04 & 0.43 $\pm$ 0.06 & 1.40 $\pm$ 0.00 & 0.29 & 7 & 8 \\ 
 99248           & 99248 & 16.30 & 0.15 & 1.14 $\pm$ 0.37 & 0.41 $\pm$ 0.28 & 1.40 $\pm$ 0.48 & 0.48 & 0 & 8 \\ 
 137099           & D7099 & 18.20 & 0.15 & 0.56 $\pm$ 0.02 & 0.29 $\pm$ 0.04 & 1.40 $\pm$ 0.00 & 0.65 & 6 & 6 \\ 
 138127           & D8127 & 17.10 & 0.15 & 0.75 $\pm$ 0.02 & 0.45 $\pm$ 0.06 & 1.40 $\pm$ 0.00 & 0.17 & 7 & 7 \\ 
 138947           & D8947 & 18.70 & 0.15 & 0.45 $\pm$ 0.12 & 0.29 $\pm$ 0.28 & 1.40 $\pm$ 0.46 & 0.46 & 0 & 9 \\ 
 142781           & E2781 & 16.10 & 0.15 & 1.59 $\pm$ 0.05 & 0.25 $\pm$ 0.04 & 1.40 $\pm$ 0.00 & 0.15 & 14 & 14 \\ 
 142781           & E2781 & 16.10 & 0.15 & 2.01 $\pm$ 0.74 & 0.16 $\pm$ 0.15 & 1.40 $\pm$ 0.44 & 0.85 & 0 & 15 \\ 
 142781           & E2781 & 16.10 & 0.15 & 2.03 $\pm$ 0.77 & 0.16 $\pm$ 0.09 & 1.40 $\pm$ 0.40 & 0.45 & 0 & 9 \\ 
 143624           & E3624 & 15.90 & 0.15 & 2.14 $\pm$ 0.04 & 0.17 $\pm$ 0.03 & 1.40 $\pm$ 0.00 & 0.32 & 9 & 9 \\ 
 143624           & E3624 & 15.90 & 0.15 & 2.23 $\pm$ 1.08 & 0.15 $\pm$ 0.17 & 1.40 $\pm$ 0.53 & 0.82 & 0 & 8 \\ 
 154276           & F4276 & 17.60 & 0.15 & 1.06 $\pm$ 0.35 & 0.14 $\pm$ 0.17 & 1.40 $\pm$ 0.43 & 0.29 & 0 & 5 \\ 
 159454           & F9454 & 17.90 & 0.15 & 0.58 $\pm$ 0.02 & 0.37 $\pm$ 0.04 & 1.40 $\pm$ 0.00 & 0.30 & 6 & 6 \\ 
 159560           & F9560 & 17.00 & 0.15 & 1.10 $\pm$ 0.47 & 0.24 $\pm$ 0.23 & 1.40 $\pm$ 0.54 & 1.16 & 0 & 87 \\ 
 159560           & F9560 & 17.00 & 0.15 & 1.16 $\pm$ 0.30 & 0.21 $\pm$ 0.21 & 1.40 $\pm$ 0.39 & 0.53 & 0 & 13 \\ 
 159857           & F9857 & 15.40 & 0.15 & 3.07 $\pm$ 1.32 & 0.13 $\pm$ 0.16 & 1.40 $\pm$ 0.45 & 0.34 & 0 & 5 \\ 
 162058           & G2058 & 17.80 & 0.15 & 0.85 $\pm$ 0.01 & 0.19 $\pm$ 0.02 & 1.40 $\pm$ 0.00 & 0.34 & 26 & 27 \\ 
 162058           & G2058 & 17.80 & 0.15 & 0.85 $\pm$ 0.28 & 0.19 $\pm$ 0.14 & 1.40 $\pm$ 0.44 & 0.87 & 0 & 31 \\ 
 162080           & G2080 & 19.80 & 0.15 & 0.78 $\pm$ 0.06 & 0.04 $\pm$ 0.01 & 1.40 $\pm$ 0.11 & 1.39 & 4 & 4 \\ 
 162080           & G2080 & 19.80 & 0.15 & 0.82 $\pm$ 0.33 & 0.03 $\pm$ 0.07 & 1.40 $\pm$ 0.49 & 0.99 & 0 & 13 \\ 
 162116           & G2116 & 19.30 & 0.15 & 0.54 $\pm$ 0.17 & 0.12 $\pm$ 0.08 & 1.40 $\pm$ 0.40 & 0.47 & 0 & 7 \\ 
 162567           & G2567 & 19.90 & 0.15 & 0.33 $\pm$ 0.01 & 0.17 $\pm$ 0.03 & 1.40 $\pm$ 0.00 & 0.20 & 6 & 6 \\ 
 162741           & G2741 & 17.30 & 0.15 & 3.95 $\pm$ 0.04 & 0.01 $\pm$ 0.00 & 1.40 $\pm$ 0.00 & 0.22 & 6 & 6 \\ 
 162980           & G2980 & 16.70 & 0.15 & 0.79 $\pm$ 0.04 & 0.66 $\pm$ 0.13 & 1.40 $\pm$ 0.00 & 0.40 & 8 & 8 \\ 
 163818           & G3818 & 18.40 & 0.15 & 0.39 $\pm$ 0.02 & 0.52 $\pm$ 0.06 & 1.40 $\pm$ 0.00 & 0.33 & 7 & 7 \\ 
 172034           & H2034 & 17.80 & 0.15 & 0.63 $\pm$ 0.02 & 0.34 $\pm$ 0.05 & 1.40 $\pm$ 0.00 & 1.05 & 16 & 16 \\ 
 190166           & J0166 & 17.10 & 0.15 & 1.01 $\pm$ 0.03 & 0.25 $\pm$ 0.04 & 1.40 $\pm$ 0.00 & 0.92 & 6 & 7 \\ 
 190166           & J0166 & 17.10 & 0.15 & 1.05 $\pm$ 0.02 & 0.23 $\pm$ 0.03 & 1.40 $\pm$ 0.00 & 0.68 & 12 & 12 \\ 
 209924           & K9924 & 16.10 & 0.15 & 1.86 $\pm$ 0.71 & 0.19 $\pm$ 0.12 & 1.40 $\pm$ 0.44 & 0.40 & 0 & 7 \\ 
 211871           & L1871 & 18.80 & 0.15 & 0.41 $\pm$ 0.01 & 0.32 $\pm$ 0.05 & 1.40 $\pm$ 0.00 & 0.28 & 5 & 7 \\ 
 214088           & L4088 & 15.20 & 0.15 & 2.42 $\pm$ 0.06 & 0.25 $\pm$ 0.03 & 1.40 $\pm$ 0.00 & 0.63 & 8 & 8 \\ 
 215588           & L5588 & 19.50 & 0.15 & 0.49 $\pm$ 0.16 & 0.12 $\pm$ 0.12 & 1.40 $\pm$ 0.44 & 0.57 & 0 & 5 \\ 
 215757           & L5757 & 17.70 & 0.15 & 0.78 $\pm$ 0.27 & 0.24 $\pm$ 0.17 & 1.40 $\pm$ 0.48 & 0.47 & 0 & 11 \\ 
 235086           & N5086 & 17.50 & 0.15 & 1.02 $\pm$ 0.40 & 0.17 $\pm$ 0.18 & 1.40 $\pm$ 0.51 & 1.04 & 0 & 60 \\ 
 235086           & N5086 & 17.50 & 0.15 & 1.02 $\pm$ 0.32 & 0.17 $\pm$ 0.11 & 1.40 $\pm$ 0.38 & 1.64 & 0 & 32 \\ 
 235086           & N5086 & 17.50 & 0.15 & 1.08 $\pm$ 0.33 & 0.15 $\pm$ 0.12 & 1.40 $\pm$ 0.38 & 0.85 & 0 & 29 \\ 
 242450           & O2450 & 14.70 & 0.15 & 2.54 $\pm$ 0.10 & 0.36 $\pm$ 0.13 & 1.40 $\pm$ 0.00 & 0.41 & 11 & 11 \\ 
 242450           & O2450 & 14.70 & 0.15 & 2.91 $\pm$ 0.08 & 0.27 $\pm$ 0.04 & 1.40 $\pm$ 0.00 & 0.83 & 13 & 14 \\ 
 250620           & P0620 & 18.00 & 0.15 & 0.65 $\pm$ 0.14 & 0.26 $\pm$ 0.13 & 1.40 $\pm$ 0.33 & 0.29 & 0 & 4 \\ 
 267337           & Q7337 & 18.00 & 0.15 & 0.44 $\pm$ 0.10 & 0.58 $\pm$ 0.25 & 1.40 $\pm$ 0.43 & 0.21 & 0 & 4 \\ 
 269690           & Q9690 & 18.40 & 0.15 & 0.89 $\pm$ 0.43 & 0.10 $\pm$ 0.11 & 1.40 $\pm$ 0.59 & 0.31 & 0 & 7 \\ 
 271480           & R1480 & 17.50 & 0.15 & 0.71 $\pm$ 0.22 & 0.35 $\pm$ 0.22 & 1.40 $\pm$ 0.48 & 0.82 & 0 & 6 \\ 
 274138           & R4138 & 17.80 & 0.15 & 0.75 $\pm$ 0.02 & 0.24 $\pm$ 0.03 & 1.40 $\pm$ 0.00 & 0.48 & 7 & 7 \\ 
 275976           & R5976 & 16.30 & 0.15 & 1.86 $\pm$ 0.04 & 0.15 $\pm$ 0.03 & 1.40 $\pm$ 0.00 & 1.01 & 5 & 5 \\ 
 275976           & R5976 & 16.30 & 0.15 & 2.38 $\pm$ 0.03 & 0.09 $\pm$ 0.01 & 1.40 $\pm$ 0.00 & 1.11 & 15 & 16 \\ 
 276274           & R6274 & 17.20 & 0.15 & 1.53 $\pm$ 0.71 & 0.10 $\pm$ 0.17 & 1.40 $\pm$ 0.52 & 0.91 & 0 & 5 \\ 
 276468           & R6468 & 17.90 & 0.15 & 1.03 $\pm$ 0.37 & 0.11 $\pm$ 0.14 & 1.40 $\pm$ 0.42 & 0.36 & 0 & 5 \\ 
 285944           & S5944 & 16.50 & 0.15 & 1.04 $\pm$ 0.04 & 0.41 $\pm$ 0.03 & 1.40 $\pm$ 0.00 & 0.16 & 10 & 10 \\ 
 285944           & S5944 & 16.50 & 0.15 & 1.40 $\pm$ 0.43 & 0.23 $\pm$ 0.19 & 1.40 $\pm$ 0.41 & 0.51 & 0 & 29 \\ 
 297418           & T7418 & 18.60 & 0.15 & 0.41 $\pm$ 0.02 & 0.39 $\pm$ 0.05 & 1.40 $\pm$ 0.00 & 0.93 & 5 & 5 \\ 
 299582           & T9582 & 18.00 & 0.15 & 0.62 $\pm$ 0.02 & 0.29 $\pm$ 0.03 & 1.40 $\pm$ 0.00 & 0.31 & 7 & 7 \\ 
 303174           & U3174 & 16.70 & 0.15 & 1.50 $\pm$ 0.03 & 0.16 $\pm$ 0.03 & 1.40 $\pm$ 0.00 & 0.65 & 21 & 23 \\ 
 304330           & U4330 & 18.90 & 0.15 & 0.61 $\pm$ 0.01 & 0.13 $\pm$ 0.02 & 1.40 $\pm$ 0.00 & 0.13 & 11 & 11 \\ 
 304330           & U4330 & 18.90 & 0.15 & 0.78 $\pm$ 0.01 & 0.08 $\pm$ 0.01 & 1.40 $\pm$ 0.00 & 0.23 & 12 & 12 \\ 
 322763           & W2763 & 16.90 & 0.15 & 1.25 $\pm$ 0.03 & 0.20 $\pm$ 0.04 & 1.40 $\pm$ 0.00 & 0.27 & 12 & 13 \\ 
 326388           & W6388 & 18.20 & 0.15 & 1.26 $\pm$ 0.57 & 0.06 $\pm$ 0.12 & 1.40 $\pm$ 0.52 & 0.33 & 0 & 8 \\ 
 334673           & X4673 & 17.90 & 0.15 & 0.57 $\pm$ 0.22 & 0.38 $\pm$ 0.25 & 1.40 $\pm$ 0.60 & 0.67 & 0 & 11 \\ 
 349219           & Y9219 & 18.20 & 0.15 & 0.58 $\pm$ 0.15 & 0.27 $\pm$ 0.23 & 1.40 $\pm$ 0.41 & 0.58 & 0 & 14 \\ 
 363505           & a3505 & 18.10 & 0.15 & 1.90 $\pm$ 0.05 & 0.03 $\pm$ 0.01 & 1.40 $\pm$ 0.03 & 0.66 & 12 & 12 \\ 
 368184           & a8184 & 19.50 & 0.15 & 0.38 $\pm$ 0.12 & 0.19 $\pm$ 0.19 & 1.40 $\pm$ 0.46 & 0.54 & 0 & 25 \\ 
 369264           & a9264 & 16.30 & 0.15 & 1.51 $\pm$ 0.47 & 0.23 $\pm$ 0.20 & 1.40 $\pm$ 0.42 & 0.65 & 0 & 7 \\ 
 377732           & b7732 & 17.00 & 0.15 & 0.95 $\pm$ 0.03 & 0.31 $\pm$ 0.05 & 1.40 $\pm$ 0.00 & 0.62 & 5 & 5 \\ 
 377732           & b7732 & 17.00 & 0.15 & 0.99 $\pm$ 0.03 & 0.29 $\pm$ 0.03 & 1.40 $\pm$ 0.00 & 0.16 & 5 & 5 \\ 
 381677           & c1677 & 18.40 & 0.15 & 0.47 $\pm$ 0.01 & 0.35 $\pm$ 0.05 & 1.40 $\pm$ 0.00 & 0.92 & 19 & 19 \\ 
 381677           & c1677 & 18.40 & 0.15 & 0.44 $\pm$ 0.16 & 0.40 $\pm$ 0.21 & 1.40 $\pm$ 0.54 & 0.47 & 0 & 5 \\ 
 387733           & c7733 & 18.90 & 0.15 & 0.34 $\pm$ 0.01 & 0.41 $\pm$ 0.06 & 1.40 $\pm$ 0.00 & 0.22 & 11 & 11 \\ 
 387733           & c7733 & 18.90 & 0.15 & 0.32 $\pm$ 0.09 & 0.47 $\pm$ 0.25 & 1.40 $\pm$ 0.46 & 0.37 & 0 & 5 \\ 
 387746           & c7746 & 20.00 & 0.15 & 0.37 $\pm$ 0.01 & 0.13 $\pm$ 0.02 & 1.40 $\pm$ 0.00 & 0.23 & 5 & 6 \\ 
 388838           & c8838 & 19.50 & 0.15 & 0.36 $\pm$ 0.01 & 0.21 $\pm$ 0.04 & 1.40 $\pm$ 0.00 & 0.61 & 18 & 18 \\ 
 388838           & c8838 & 19.50 & 0.15 & 0.38 $\pm$ 0.01 & 0.20 $\pm$ 0.02 & 1.40 $\pm$ 0.00 & 0.24 & 12 & 12 \\ 
 389694           & c9694 & 18.20 & 0.15 & 0.45 $\pm$ 0.02 & 0.46 $\pm$ 0.06 & 1.40 $\pm$ 0.00 & 0.22 & 4 & 5 \\ 
 391211           & d1211 & 18.50 & 0.15 & 0.41 $\pm$ 0.09 & 0.42 $\pm$ 0.23 & 1.40 $\pm$ 0.38 & 0.85 & 0 & 17 \\ 
 393359           & d3359 & 19.20 & 0.15 & 0.77 $\pm$ 0.33 & 0.06 $\pm$ 0.11 & 1.40 $\pm$ 0.52 & 0.51 & 0 & 30 \\ 
 393569           & d3569 & 20.20 & 0.15 & 0.55 $\pm$ 0.01 & 0.05 $\pm$ 0.01 & 1.40 $\pm$ 0.00 & 0.22 & 13 & 14 \\ 
 399433           & d9433 & 18.60 & 0.15 & 1.34 $\pm$ 0.56 & 0.04 $\pm$ 0.09 & 1.40 $\pm$ 0.49 & 0.24 & 0 & 10 \\ 
 399433           & d9433 & 18.60 & 0.15 & 1.76 $\pm$ 0.89 & 0.02 $\pm$ 0.05 & 1.40 $\pm$ 0.53 & 0.16 & 0 & 9 \\ 
 406952           & e6952 & 17.10 & 0.15 & 0.77 $\pm$ 0.21 & 0.43 $\pm$ 0.23 & 1.40 $\pm$ 0.44 & 0.64 & 0 & 5 \\ 
 408751           & e8751 & 19.00 & 0.15 & 0.40 $\pm$ 0.01 & 0.28 $\pm$ 0.03 & 1.40 $\pm$ 0.00 & 0.84 & 68 & 69 \\ 
 409256           & e9256 & 18.20 & 0.15 & 1.89 $\pm$ 0.68 & 0.03 $\pm$ 0.04 & 1.40 $\pm$ 0.40 & 0.84 & 0 & 4 \\ 
 409836           & e9836 & 18.10 & 0.15 & 0.55 $\pm$ 0.19 & 0.33 $\pm$ 0.25 & 1.40 $\pm$ 0.49 & 1.78 & 0 & 14 \\ 
 410088           & f0088 & 18.10 & 0.15 & 1.03 $\pm$ 0.01 & 0.10 $\pm$ 0.02 & 1.40 $\pm$ 0.00 & 0.14 & 9 & 10 \\ 
 410778           & f0778 & 18.10 & 0.15 & 1.46 $\pm$ 0.57 & 0.05 $\pm$ 0.03 & 1.40 $\pm$ 0.41 & 0.38 & 0 & 6 \\ 
 411201           & f1201 & 17.80 & 0.15 & 0.66 $\pm$ 0.01 & 0.31 $\pm$ 0.05 & 1.40 $\pm$ 0.00 & 1.45 & 12 & 15 \\ 
 411611           & f1611 & 18.80 & 0.15 & 0.36 $\pm$ 0.10 & 0.41 $\pm$ 0.21 & 1.40 $\pm$ 0.43 & 0.81 & 0 & 31 \\ 
 413038           & f3038 & 16.90 & 0.15 & 1.24 $\pm$ 0.03 & 0.20 $\pm$ 0.04 & 1.40 $\pm$ 0.00 & 1.19 & 22 & 23 \\ 
 413038           & f3038 & 16.90 & 0.15 & 1.01 $\pm$ 0.04 & 0.30 $\pm$ 0.04 & 1.40 $\pm$ 0.00 & 1.79 & 23 & 25 \\ 
 413192           & f3192 & 16.80 & 0.15 & 3.96 $\pm$ 1.84 & 0.02 $\pm$ 0.05 & 1.40 $\pm$ 0.47 & 0.62 & 0 & 18 \\ 
 413421           & f3421 & 18.30 & 0.15 & 1.90 $\pm$ 0.78 & 0.02 $\pm$ 0.02 & 1.40 $\pm$ 0.41 & 1.39 & 0 & 23 \\ 
 413820           & f3820 & 19.80 & 0.15 & 0.66 $\pm$ 0.26 & 0.05 $\pm$ 0.04 & 1.40 $\pm$ 0.46 & 0.97 & 0 & 36 \\ 
 414286           & f4286 & 18.60 & 0.15 & 0.37 $\pm$ 0.08 & 0.47 $\pm$ 0.19 & 1.40 $\pm$ 0.38 & 0.54 & 0 & 27 \\ 
 414286           & f4286 & 18.60 & 0.15 & 0.40 $\pm$ 0.09 & 0.40 $\pm$ 0.24 & 1.40 $\pm$ 0.40 & 0.71 & 0 & 29 \\ 
 418797           & f8797 & 19.40 & 0.15 & 0.70 $\pm$ 0.29 & 0.06 $\pm$ 0.07 & 1.40 $\pm$ 0.50 & 0.32 & 0 & 7 \\ 
 418929           & f8929 & 17.00 & 0.15 & 1.43 $\pm$ 0.02 & 0.14 $\pm$ 0.03 & 1.40 $\pm$ 0.00 & 0.54 & 48 & 49 \\ 
 419624           & f9624 & 20.50 & 0.15 & 0.34 $\pm$ 0.14 & 0.09 $\pm$ 0.17 & 1.40 $\pm$ 0.50 & 0.57 & 0 & 18 \\ 
 419624           & f9624 & 20.50 & 0.15 & 0.36 $\pm$ 0.13 & 0.09 $\pm$ 0.14 & 1.40 $\pm$ 0.46 & 0.46 & 0 & 6 \\ 
 419880           & f9880 & 19.60 & 0.15 & 0.98 $\pm$ 0.06 & 0.03 $\pm$ 0.01 & 1.40 $\pm$ 0.08 & 0.20 & 6 & 6 \\ 
 2000 AG205           & K00AK5G & 19.70 & 0.15 & 0.95 $\pm$ 0.01 & 0.03 $\pm$ 0.00 & 1.40 $\pm$ 0.00 & 0.79 & 12 & 14 \\ 
 2002  XS40           & K02X40S & 20.10 & 0.15 & 0.76 $\pm$ 0.03 & 0.03 $\pm$ 0.00 & 1.40 $\pm$ 0.05 & 0.18 & 14 & 14 \\ 
 2003  CC11           & K03C11C & 19.10 & 0.15 & 1.13 $\pm$ 0.51 & 0.03 $\pm$ 0.10 & 1.40 $\pm$ 0.53 & 0.49 & 0 & 16 \\ 
 2003 SS214           & K03SL4S & 20.10 & 0.15 & 0.86 $\pm$ 0.25 & 0.02 $\pm$ 0.02 & 1.40 $\pm$ 0.35 & 0.75 & 0 & 16 \\ 
 2004  BZ74           & K04B74Z & 18.10 & 0.15 & 0.96 $\pm$ 0.02 & 0.11 $\pm$ 0.02 & 1.40 $\pm$ 0.00 & 0.65 & 4 & 5 \\ 
 2004   MX2           & K04M02X & 19.30 & 0.15 & 1.26 $\pm$ 0.08 & 0.02 $\pm$ 0.00 & 1.40 $\pm$ 0.09 & 0.35 & 9 & 9 \\ 
 2004  TG10           & K04T10G & 19.40 & 0.15 & 1.32 $\pm$ 0.61 & 0.02 $\pm$ 0.04 & 1.40 $\pm$ 0.51 & 0.64 & 0 & 8 \\ 
 2005   LS3           & K05L03S & 19.50 & 0.15 & 0.38 $\pm$ 0.10 & 0.19 $\pm$ 0.12 & 1.40 $\pm$ 0.38 & 0.64 & 0 & 7 \\ 
 2006  BB27           & K06B27B & 20.00 & 0.15 & 0.22 $\pm$ 0.05 & 0.38 $\pm$ 0.21 & 1.40 $\pm$ 0.38 & 0.98 & 0 & 5 \\ 
 2007    BG           & K07B00G & 19.50 & 0.15 & 0.31 $\pm$ 0.11 & 0.24 $\pm$ 0.19 & 1.40 $\pm$ 0.51 & 0.38 & 3 & 5 \\ 
 2007  RU10           & K07R10U & 19.10 & 0.15 & 0.92 $\pm$ 0.37 & 0.05 $\pm$ 0.06 & 1.40 $\pm$ 0.47 & 0.31 & 0 & 9 \\ 
 2008  QS11           & K08Q11S & 19.90 & 0.15 & 0.45 $\pm$ 0.01 & 0.09 $\pm$ 0.01 & 1.40 $\pm$ 0.00 & 0.33 & 9 & 11 \\ 
 2009   ND1           & K09N01D & 17.10 & 0.15 & 2.50 $\pm$ 0.95 & 0.04 $\pm$ 0.04 & 1.40 $\pm$ 0.39 & 0.70 & 0 & 11 \\ 
 2010   OQ1           & K10O01Q & 19.00 & 0.15 & 0.54 $\pm$ 0.21 & 0.15 $\pm$ 0.14 & 1.40 $\pm$ 0.51 & 0.47 & 0 & 8 \\ 
 2011   CQ4           & K11C04Q & 18.40 & 0.15 & 0.66 $\pm$ 0.02 & 0.18 $\pm$ 0.02 & 1.40 $\pm$ 0.00 & 0.29 & 5 & 7 \\ 
 2012    DN           & K12D00N & 18.10 & 0.15 & 2.77 $\pm$ 1.05 & 0.01 $\pm$ 0.03 & 1.40 $\pm$ 0.38 & 0.42 & 0 & 7 \\ 
 2013   PX6           & K13P06X & 18.40 & 0.15 & 1.65 $\pm$ 0.03 & 0.03 $\pm$ 0.00 & 1.40 $\pm$ 0.02 & 0.19 & 9 & 10 \\ 
 2013  WT44           & K13W44T & 19.30 & 0.15 & 0.65 $\pm$ 0.01 & 0.08 $\pm$ 0.02 & 1.40 $\pm$ 0.00 & 0.31 & 6 & 6 \\ 
 2013  WU44           & K13W44U & 21.00 & 0.15 & 0.29 $\pm$ 0.13 & 0.09 $\pm$ 0.19 & 1.40 $\pm$ 0.61 & 0.26 & 0 & 8 \\ 
 2013  YZ13           & K13Y13Z & 19.60 & 0.15 & 0.31 $\pm$ 0.10 & 0.27 $\pm$ 0.19 & 1.40 $\pm$ 0.46 & 0.07 & 0 & 6 \\ 
 2013 YP139           & K13YD9P & 21.60 & 0.15 & 0.40 $\pm$ 0.03 & 0.03 $\pm$ 0.01 & 1.09 $\pm$ 0.07 & 0.25 & 6 & 6 \\ 
 2014  AA33           & K14A33A & 19.30 & 0.15 & 0.79 $\pm$ 0.04 & 0.05 $\pm$ 0.01 & 1.40 $\pm$ 0.06 & 0.19 & 4 & 4 \\ 
 2014  AQ46           & K14A46Q & 20.10 & 0.15 & 0.59 $\pm$ 0.29 & 0.05 $\pm$ 0.11 & 1.40 $\pm$ 0.60 & 0.47 & 0 & 17 \\ 
 2014  AA53           & K14A53A & 19.80 & 0.15 & 0.70 $\pm$ 0.27 & 0.04 $\pm$ 0.06 & 1.40 $\pm$ 0.47 & 0.50 & 0 & 13 \\ 
 2014  BG60           & K14B60G & 20.10 & 0.15 & 0.67 $\pm$ 0.25 & 0.04 $\pm$ 0.08 & 1.40 $\pm$ 0.46 & 1.30 & 0 & 163 \\ 
 2014  BE63           & K14B63E & 23.20 & 0.15 & 0.36 $\pm$ 0.13 & 0.01 $\pm$ 0.00 & 1.40 $\pm$ 0.46 & 0.42 & 0 & 5 \\ 
 2014   CY4           & K14C04Y & 21.10 & 0.15 & 0.57 $\pm$ 0.25 & 0.02 $\pm$ 0.04 & 1.40 $\pm$ 0.52 & 0.35 & 0 & 5 \\ 
 2014  DC10           & K14D10C & 20.10 & 0.15 & 0.89 $\pm$ 0.01 & 0.02 $\pm$ 0.00 & 1.40 $\pm$ 0.00 & 0.90 & 9 & 10 \\ 
 2014    ED           & K14E00D & 19.30 & 0.15 & 0.49 $\pm$ 0.13 & 0.14 $\pm$ 0.14 & 1.40 $\pm$ 0.39 & 0.57 & 0 & 6 \\ 
 2014  EN45           & K14E45N & 21.20 & 0.15 & 0.37 $\pm$ 0.13 & 0.04 $\pm$ 0.01 & 0.75 $\pm$ 0.24 & 0.16 & 12 & 12 \\ 
 2014  EZ48           & K14E48Z & 18.80 & 0.15 & 0.45 $\pm$ 0.01 & 0.26 $\pm$ 0.04 & 1.40 $\pm$ 0.00 & 1.10 & 5 & 6 \\ 
 2014  EZ48           & K14E48Z & 18.80 & 0.15 & 0.44 $\pm$ 0.11 & 0.27 $\pm$ 0.21 & 1.40 $\pm$ 0.38 & 0.47 & 0 & 6 \\ 
 2014  EQ49           & K14E49Q & 21.80 & 0.15 & 0.38 $\pm$ 0.13 & 0.02 $\pm$ 0.03 & 1.40 $\pm$ 0.42 & 0.42 & 0 & 5 \\ 
 2014  ER49           & K14E49R & 18.60 & 0.15 & 0.46 $\pm$ 0.15 & 0.30 $\pm$ 0.26 & 1.40 $\pm$ 0.49 & 0.51 & 0 & 9 \\ 
 2014   HE3           & K14H03E & 19.90 & 0.15 & 0.56 $\pm$ 0.15 & 0.06 $\pm$ 0.04 & 1.40 $\pm$ 0.34 & 0.18 & 0 & 5 \\ 
 2014 HQ124           & K14HC4Q & 18.90 & 0.15 & 0.41 $\pm$ 0.17 & 0.29 $\pm$ 0.22 & 1.40 $\pm$ 0.57 & 0.80 & 0 & 10 \\ 
 2014 HF177           & K14HH7F & 19.70 & 0.15 & 0.25 $\pm$ 0.01 & 0.36 $\pm$ 0.06 & 1.40 $\pm$ 0.00 & 0.39 & 10 & 12 \\ 
 2014  JL25           & K14J25L & 23.00 & 0.15 & 0.23 $\pm$ 0.06 & 0.02 $\pm$ 0.03 & 1.40 $\pm$ 0.34 & 0.68 & 0 & 5 \\ 
 2014  JH57           & K14J57H & 16.60 & 0.15 & 4.61 $\pm$ 0.03 & 0.02 $\pm$ 0.00 & 1.40 $\pm$ 0.00 & 0.11 & 6 & 6 \\ 
 2014  JH57           & K14J57H & 16.60 & 0.15 & 6.79 $\pm$ 3.81 & 0.01 $\pm$ 0.03 & 1.40 $\pm$ 0.47 & 0.30 & 0 & 5 \\ 
 2014  JN57           & K14J57N & 20.70 & 0.15 & 0.27 $\pm$ 0.10 & 0.12 $\pm$ 0.10 & 1.40 $\pm$ 0.47 & 0.69 & 0 & 4 \\ 
 2014  KX99           & K14K99X & 18.20 & 0.15 & 1.72 $\pm$ 0.68 & 0.03 $\pm$ 0.05 & 1.40 $\pm$ 0.46 & 0.43 & 0 & 9 \\ 
 2014  LQ25           & K14L25Q & 20.00 & 0.15 & 0.94 $\pm$ 0.32 & 0.02 $\pm$ 0.01 & 1.40 $\pm$ 0.37 & 0.48 & 0 & 5 \\ 
 2014  LR26           & K14L26R & 18.50 & 0.15 & 2.08 $\pm$ 0.90 & 0.02 $\pm$ 0.03 & 1.40 $\pm$ 0.46 & 0.65 & 0 & 6 \\ 
 2014  MQ18           & K14M18Q & 15.60 & 0.15 & 5.27 $\pm$ 3.50 & 0.04 $\pm$ 0.07 & 1.40 $\pm$ 0.52 & 0.54 & 0 & 8 \\ 
 2014  NB39           & K14N39B & 19.50 & 0.15 & 1.08 $\pm$ 0.15 & 0.02 $\pm$ 0.02 & 1.40 $\pm$ 0.18 & 0.08 & 7 & 7 \\ 
 2014  NE52           & K14N52E & 17.90 & 0.15 & 0.70 $\pm$ 0.22 & 0.25 $\pm$ 0.27 & 1.40 $\pm$ 0.47 & 0.66 & 0 & 9 \\ 
 2014  NC64           & K14N64C & 20.50 & 0.15 & 0.50 $\pm$ 0.19 & 0.04 $\pm$ 0.02 & 0.82 $\pm$ 0.29 & 0.64 & 5 & 6 \\ 
 2014  NM64           & K14N64M & 22.60 & 0.15 & 0.33 $\pm$ 0.12 & 0.01 $\pm$ 0.02 & 1.40 $\pm$ 0.44 & 0.82 & 0 & 25 \\ 
 2014   OY1           & K14O01Y & 19.10 & 0.15 & 0.60 $\pm$ 0.21 & 0.11 $\pm$ 0.09 & 1.40 $\pm$ 0.43 & 0.30 & 0 & 6 \\ 
 2014   OZ1           & K14O01Z & 21.00 & 0.15 & 0.73 $\pm$ 0.29 & 0.01 $\pm$ 0.03 & 1.40 $\pm$ 0.49 & 0.38 & 0 & 21 \\ 
 2014  PC68           & K14P68C & 20.40 & 0.15 & 0.56 $\pm$ 0.20 & 0.04 $\pm$ 0.04 & 1.40 $\pm$ 0.43 & 0.39 & 0 & 8 \\ 
 2014  PF68           & K14P68F & 18.20 & 0.15 & 3.33 $\pm$ 2.06 & 0.01 $\pm$ 0.01 & 1.20 $\pm$ 0.48 & 0.60 & 0 & 12 \\ 
 2014 QK433           & K14Qh3K & 18.30 & 0.15 & 1.78 $\pm$ 0.75 & 0.03 $\pm$ 0.06 & 1.40 $\pm$ 0.47 & 0.79 & 0 & 10 \\ 
 2014  RH12           & K14R12H & 23.50 & 0.15 & 0.09 $\pm$ 0.04 & 0.09 $\pm$ 0.11 & 1.40 $\pm$ 0.54 & 0.75 & 0 & 10 \\ 
 2014  RL12           & K14R12L & 17.90 & 0.15 & 0.69 $\pm$ 0.02 & 0.25 $\pm$ 0.03 & 1.40 $\pm$ 0.00 & 0.31 & 5 & 5 \\ 
 2014  RL12           & K14R12L & 17.90 & 0.15 & 0.61 $\pm$ 0.17 & 0.33 $\pm$ 0.19 & 1.40 $\pm$ 0.42 & 0.83 & 0 & 6 \\ 
 2014 SR339           & K14SX9R & 18.60 & 0.15 & 0.97 $\pm$ 0.37 & 0.07 $\pm$ 0.07 & 1.40 $\pm$ 0.46 & 0.69 & 0 & 13 \\ 
 2014  TW57           & K14T57W & 20.10 & 0.15 & 0.47 $\pm$ 0.01 & 0.07 $\pm$ 0.02 & 1.40 $\pm$ 0.00 & 0.76 & 4 & 6 \\ 
 2014  TF64           & K14T64F & 20.10 & 0.15 & 0.70 $\pm$ 0.20 & 0.03 $\pm$ 0.03 & 1.40 $\pm$ 0.35 & 0.33 & 0 & 5 \\ 
 2014  TJ64           & K14T64J & 21.30 & 0.15 & 0.52 $\pm$ 0.20 & 0.02 $\pm$ 0.02 & 1.40 $\pm$ 0.47 & 0.46 & 0 & 31 \\ 
 2014  TJ64           & K14T64J & 21.30 & 0.15 & 0.52 $\pm$ 0.23 & 0.02 $\pm$ 0.03 & 1.40 $\pm$ 0.54 & 0.55 & 0 & 14 \\ 
 2014 UG176           & K14UH6G & 21.50 & 0.15 & 0.42 $\pm$ 0.12 & 0.03 $\pm$ 0.03 & 1.40 $\pm$ 0.39 & 0.17 & 0 & 8 \\ 
 2014 US192           & K14UJ2S & 18.70 & 0.15 & 0.87 $\pm$ 0.01 & 0.08 $\pm$ 0.01 & 1.40 $\pm$ 0.00 & 0.25 & 5 & 5 \\ 
 2014 UF206           & K14UK6F & 18.80 & 0.15 & 1.63 $\pm$ 0.79 & 0.02 $\pm$ 0.04 & 1.40 $\pm$ 0.49 & 0.62 & 0 & 17 \\ 
 2014 UH210           & K14UL0H & 21.10 & 0.15 & 0.40 $\pm$ 0.16 & 0.04 $\pm$ 0.06 & 1.40 $\pm$ 0.47 & 0.76 & 0 & 5 \\ 
 2014  VP35           & K14V35P & 22.70 & 0.15 & 0.12 $\pm$ 0.05 & 0.10 $\pm$ 0.10 & 1.40 $\pm$ 0.53 & 0.36 & 0 & 6 \\ 
 2014  WJ70           & K14W70J & 17.60 & 0.15 & 2.92 $\pm$ 1.21 & 0.02 $\pm$ 0.04 & 1.40 $\pm$ 0.44 & 0.62 & 0 & 27 \\ 
 2014   XQ7           & K14X07Q & 20.60 & 0.15 & 0.65 $\pm$ 0.29 & 0.02 $\pm$ 0.05 & 1.40 $\pm$ 0.55 & 0.83 & 0 & 8 \\ 
 2014   XX7           & K14X07X & 19.80 & 0.15 & 1.20 $\pm$ 0.38 & 0.01 $\pm$ 0.02 & 1.40 $\pm$ 0.36 & 0.43 & 0 & 6 \\ 
 2014  XX31           & K14X31X & 17.60 & 0.15 & 1.35 $\pm$ 0.49 & 0.09 $\pm$ 0.15 & 1.40 $\pm$ 0.43 & 0.42 & 0 & 8 \\ 
\enddata
\label{tab:neonew}
\end{deluxetable}
 
\begin{deluxetable}{rrrrrrrrrr}
\tabletypesize{\scriptsize}

\tablecaption{Measured diameters ($d$) and albedos ($p_V$) of near-Earth asteroids. Objects in this table have previously reported measurements by the NEOWISE team \citep{Mainzer2011d,Mainzer12}. Previous measurements use detections in 12 $\mu$m and 22 $\mu$m bands, and therefore are better constrained.  Magnitude $H$, slope parameter $G$, and beaming $\eta$ used are given. The numbers of observations used in the 3.4 $\mu$m ($n_{W1}$) and 4.6 $\mu$m ($n_{W2}$) wavelengths are also reported, along with the amplitude of the 4.6 $\mu$m light curve (W2 amp.).}
\tablewidth{0pt}
\tablehead{
	\colhead{Name} & \colhead{Packed} & \colhead{$H$}& \colhead{$G$}& \colhead{$d$ (km)} & \colhead{$p_V$} & \colhead{$\eta$} & \colhead{W2 amp.} & \colhead{$n_{W1}$} & \colhead{$n_{W2}$} }                                                                                   
\startdata
2102           & 02102 & 16.00 & 0.15 & 1.68 $\pm$ 0.05 & 0.25 $\pm$ 0.04 & 1.40 $\pm$ 0.00 & 0.23 & 13 & 13 \\ 
 2102           & 02102 & 16.00 & 0.15 & 1.65 $\pm$ 0.05 & 0.26 $\pm$ 0.04 & 1.40 $\pm$ 0.00 & 0.18 & 5 & 5 \\ 
 2102           & 02102 & 16.00 & 0.15 & 1.69 $\pm$ 0.06 & 0.25 $\pm$ 0.03 & 1.40 $\pm$ 0.00 & 0.67 & 8 & 9 \\ 
 3554           & 03554 & 15.82 & 0.15 & 1.56 $\pm$ 0.07 & 0.34 $\pm$ 0.06 & 1.40 $\pm$ 0.00 & 0.49 & 19 & 20 \\ 
 4183           & 04183 & 14.40 & 0.15 & 2.94 $\pm$ 0.12 & 0.36 $\pm$ 0.06 & 1.40 $\pm$ 0.00 & 1.04 & 12 & 12 \\ 
 4183           & 04183 & 14.40 & 0.15 & 3.54 $\pm$ 0.12 & 0.24 $\pm$ 0.04 & 1.40 $\pm$ 0.00 & 0.62 & 17 & 18 \\ 
 6050           & 06050 & 14.80 & 0.15 & 2.88 $\pm$ 0.07 & 0.26 $\pm$ 0.04 & 1.40 $\pm$ 0.00 & 1.51 & 57 & 57 \\ 
 25916           & 25916 & 13.60 & 0.15 & 5.96 $\pm$ 0.13 & 0.18 $\pm$ 0.03 & 1.40 $\pm$ 0.00 & 0.63 & 24 & 29 \\ 
 27346           & 27346 & 15.90 & 0.15 & 1.80 $\pm$ 0.07 & 0.24 $\pm$ 0.04 & 1.40 $\pm$ 0.00 & 0.43 & 9 & 9 \\ 
 40263           & 40263 & 17.70 & 0.15 & 0.92 $\pm$ 0.35 & 0.17 $\pm$ 0.18 & 1.40 $\pm$ 0.48 & 0.71 & 0 & 14 \\ 
 40267           & 40267 & 15.40 & 0.15 & 2.39 $\pm$ 0.09 & 0.21 $\pm$ 0.04 & 1.40 $\pm$ 0.00 & 1.06 & 4 & 4 \\ 
 85628           & 85628 & 17.00 & 0.15 & 0.78 $\pm$ 0.03 & 0.46 $\pm$ 0.08 & 1.40 $\pm$ 0.00 & 0.64 & 7 & 10 \\ 
 90367           & 90367 & 17.70 & 0.15 & 1.76 $\pm$ 0.79 & 0.05 $\pm$ 0.13 & 1.40 $\pm$ 0.51 & 0.54 & 0 & 12 \\ 
 90367           & 90367 & 17.70 & 0.15 & 2.00 $\pm$ 0.89 & 0.04 $\pm$ 0.03 & 1.40 $\pm$ 0.46 & 0.49 & 0 & 13 \\ 
 137062           & D7062 & 16.60 & 0.15 & 0.99 $\pm$ 0.06 & 0.41 $\pm$ 0.05 & 1.40 $\pm$ 0.00 & 0.89 & 6 & 6 \\ 
 138847           & D8847 & 16.90 & 0.15 & 0.94 $\pm$ 0.28 & 0.35 $\pm$ 0.19 & 1.40 $\pm$ 0.44 & 1.01 & 0 & 26 \\ 
 162181           & G2181 & 18.20 & 0.15 & 0.73 $\pm$ 0.02 & 0.17 $\pm$ 0.03 & 1.40 $\pm$ 0.00 & 0.32 & 25 & 25 \\ 
 162483           & G2483 & 17.50 & 0.15 & 0.69 $\pm$ 0.20 & 0.37 $\pm$ 0.21 & 1.40 $\pm$ 0.44 & 0.62 & 0 & 9 \\ 
 162566           & G2566 & 15.70 & 0.15 & 6.00 $\pm$ 2.42 & 0.03 $\pm$ 0.04 & 1.40 $\pm$ 0.40 & 1.02 & 0 & 24 \\ 
 163691           & G3691 & 17.00 & 0.15 & 3.06 $\pm$ 1.55 & 0.03 $\pm$ 0.06 & 1.40 $\pm$ 0.54 & 0.30 & 0 & 5 \\ 
 243566           & O3566 & 17.40 & 0.15 & 0.88 $\pm$ 0.02 & 0.25 $\pm$ 0.04 & 1.40 $\pm$ 0.00 & 0.29 & 11 & 11 \\ 
 262623           & Q2623 & 18.50 & 0.15 & 0.49 $\pm$ 0.15 & 0.29 $\pm$ 0.18 & 1.40 $\pm$ 0.44 & 0.48 & 0 & 4 \\ 
 276049           & R6049 & 16.80 & 0.15 & 4.03 $\pm$ 1.85 & 0.02 $\pm$ 0.04 & 1.40 $\pm$ 0.44 & 0.54 & 0 & 6 \\ 
 277616           & R7616 & 17.40 & 0.15 & 1.28 $\pm$ 0.01 & 0.12 $\pm$ 0.02 & 1.40 $\pm$ 0.00 & 0.28 & 4 & 4 \\ 
 395207           & d5207 & 19.60 & 0.15 & 0.60 $\pm$ 0.20 & 0.07 $\pm$ 0.03 & 1.40 $\pm$ 0.40 & 0.32 & 0 & 8 \\ 
 395207           & d5207 & 19.60 & 0.15 & 0.73 $\pm$ 0.30 & 0.05 $\pm$ 0.10 & 1.40 $\pm$ 0.49 & 0.50 & 0 & 19 \\ 
 397237           & d7237 & 16.70 & 0.15 & 1.73 $\pm$ 0.66 & 0.12 $\pm$ 0.16 & 1.40 $\pm$ 0.46 & 0.40 & 0 & 4 \\ 
  1998  SB15           & J98S15B & 20.90 & 0.15 & 0.36 $\pm$ 0.12 & 0.06 $\pm$ 0.09 & 1.40 $\pm$ 0.44 & 0.66 & 0 & 11 \\ 
 2009  UX17           & K09U17X & 21.50 & 0.15 & 0.39 $\pm$ 0.13 & 0.03 $\pm$ 0.03 & 1.40 $\pm$ 0.40 & 0.86 & 0 & 15 \\ 
 2010  LF86           & K10L86F & 17.30 & 0.15 & 2.30 $\pm$ 0.89 & 0.04 $\pm$ 0.04 & 1.40 $\pm$ 0.41 & 0.21 & 0 & 7 \\ 
 2010  LO97           & K10L97O & 18.70 & 0.15 & 1.40 $\pm$ 0.57 & 0.03 $\pm$ 0.06 & 1.40 $\pm$ 0.47 & 0.57 & 0 & 15 \\ 
 2010   NG3           & K10N03G & 17.20 & 0.15 & 1.45 $\pm$ 0.02 & 0.11 $\pm$ 0.02 & 1.40 $\pm$ 0.00 & 0.64 & 17 & 17 \\ 
 2010   NG3           & K10N03G & 17.20 & 0.15 & 1.74 $\pm$ 0.94 & 0.08 $\pm$ 0.18 & 1.40 $\pm$ 0.59 & 0.80 & 0 & 17 \\ 
 2014 HJ129           & K14HC9J & 21.10 & 0.15 & 0.59 $\pm$ 0.21 & 0.02 $\pm$ 0.02 & 1.40 $\pm$ 0.42 & 0.50 & 0 & 9 \\ 
   \enddata
 \label{tab:neodup}
 \end{deluxetable}
\begin{deluxetable}{rrrrrrrrrr}
\tabletypesize{\scriptsize}
\tablecaption{Measured diameters ($d$) and albedos ($p_V$) of MBAs and Mars crossers. Objects in this table do not have previously published diameters and albedos by the NEOWISE team. Beaming $\eta$, $H$, $G$, the amplitude of the 4.6 $\mu$m light curve (W2 amp.), and the numbers of observations used in the 3.4 $\mu$m ($n_{W1}$) and 4.6 $\mu$m ($n_{W2}$) wavelengths are also reported. For a small ($<1\%$) fraction of objects, diameter fits could not reproduce optical magnitudes for a realistic range of albedos. This may be due to a large light curve amplitude, uncertainty in $G$ slope values used to derive $H$ magnitudes, or other reasons noted in \citet{Mainzer2011d,Masiero11,Mainzer12,Masiero12}. These objects are marked with a $\dagger$ in the name column. Objects without reported albedos did not have measured $H$ values, see text for details. Only the first 15 lines are shown; the remainder are available in electronic format through the journal website. }
\tablewidth{0pt}
\tablehead{
	\colhead{Name} & \colhead{Packed} & \colhead{$H$}& \colhead{$G$}& \colhead{$d$ (km)} & \colhead{$p_V$} & \colhead{$\eta$} & \colhead{W2 amp.} & \colhead{$n_{W1}$} & \colhead{$n_{W2}$} }                                                                                        
\startdata
 21           & 00021 & 7.35 & 0.11 & 99.47 $\pm$ 27.12 & 0.16 $\pm$ 0.12 & 0.95 $\pm$ 0.19 & 0.27 & 10 & 10 \\ 
 65           & 00065 & 6.62 & 0.01 & 276.58 $\pm$ 74.49 & 0.06 $\pm$ 0.04 & 0.95 $\pm$ 0.17 & 0.09 & 8 & 10 \\ 
 69           & 00069 & 7.05 & 0.19 & 131.07 $\pm$ 32.19 & 0.19 $\pm$ 0.07 & 0.95 $\pm$ 0.18 & 0.09 & 14 & 14 \\ 
 74           & 00074 & 8.66 & 0.15 & 111.87 $\pm$ 46.38 & 0.04 $\pm$ 0.03 & 0.95 $\pm$ 0.23 & 0.26 & 3 & 4 \\ 
 74           & 00074 & 8.66 & 0.15 & 105.13 $\pm$ 29.95 & 0.05 $\pm$ 0.02 & 0.95 $\pm$ 0.16 & 0.24 & 9 & 9 \\ 
 140           & 00140 & 8.34 & 0.15 & 82.63 $\pm$ 20.19 & 0.09 $\pm$ 0.07 & 0.95 $\pm$ 0.18 & 0.37 & 7 & 7 \\ 
 144           & 00144 & 7.91 & 0.17 & 131.36 $\pm$ 33.30 & 0.05 $\pm$ 0.01 & 0.95 $\pm$ 0.17 & 0.31 & 10 & 10 \\ 
 147           & 00147 & 8.70 & 0.15 & 144.68 $\pm$ 47.63 & 0.03 $\pm$ 0.02 & 0.95 $\pm$ 0.19 & 0.11 & 6 & 6 \\ 
 147           & 00147 & 8.70 & 0.15 & 119.59 $\pm$ 37.39 & 0.04 $\pm$ 0.02 & 0.95 $\pm$ 0.18 & 0.20 & 9 & 9 \\ 
 160           & 00160 & 9.08 & 0.15 & 69.62 $\pm$ 13.23 & 0.07 $\pm$ 0.04 & 0.95 $\pm$ 0.14 & 0.58 & 20 & 21 \\ 
 212           & 00212 & 8.28 & 0.15 & 132.58 $\pm$ 48.48 & 0.05 $\pm$ 0.03 & 0.95 $\pm$ 0.20 & 0.16 & 5 & 5 \\ 
 212           & 00212 & 8.28 & 0.15 & 129.09 $\pm$ 40.48 & 0.05 $\pm$ 0.04 & 0.95 $\pm$ 0.19 & 0.17 & 7 & 7 \\ 
 253           & 00253 & 10.30 & 0.15 & 50.35 $\pm$ 17.16 & 0.04 $\pm$ 0.02 & 0.95 $\pm$ 0.24 & 0.43 & 16 & 16 \\ 
 284           & 00284 & 10.05 & 0.11 & 54.47 $\pm$ 20.59 & 0.04 $\pm$ 0.03 & 0.95 $\pm$ 0.23 & 0.21 & 11 & 11 \\ 

 \enddata
 \label{tab:mbanew}
 \end{deluxetable}

\begin{deluxetable}{rrrrrrrrrr}
\tabletypesize{\scriptsize}

\tablecaption{Measured diameters ($d$) and albedos ($p_V$) of MBA and Mars crossing asteroids. Objects in this table have previously reported measurements by the NEOWISE team \citep{Masiero11,Masiero12}. Previous measurements use detections in 12 $\mu$m and 22 $\mu$m bands, and therefore are better constrained. $H$, $G$, beaming $\eta$, the amplitude of the 4.6 $\mu$m light curve (W2 amp.), and the numbers of observations used in the 3.4 $\mu$m ($n_{W1}$) and 4.6 $\mu$m ($n_{W2}$) wavelengths are also reported.
For a small ($<1\%$) fraction of objects, diameter fits could not reproduce optical magnitudes for a realistic range of albedos. This may be due to a large light curve amplitude, uncertainty in $G$ slope values used to derive $H$ magnitudes, or other reasons noted in \citet{Mainzer2011d,Masiero11,Mainzer12,Masiero12}. These objects are marked with a $\dagger$ in the name column. Only the first 15 lines are shown; the remainder are available in electronic format through the journal website.}
\tablewidth{0pt}
\tablehead{
	\colhead{Name} & \colhead{Packed} & \colhead{$H$}& \colhead{$G$}& \colhead{$d$ (km)} & \colhead{$p_V$} & \colhead{$\eta$} & \colhead{W2 amp.} & \colhead{$n_{W1}$} & \colhead{$n_{W2}$} }                                                                                        
\startdata 
 13           & 00013 & 6.74 & 0.15 & 202.64 $\pm$ 50.08 & 0.06 $\pm$ 0.03 & 0.95 $\pm$ 0.16 & 0.32 & 9 & 9 \\ 
 24           & 00024 & 7.08 & 0.19 & 151.82 $\pm$ 49.32 & 0.08 $\pm$ 0.04 & 0.95 $\pm$ 0.20 & 0.20 & 15 & 15 \\ 
 30           & 00030 & 7.57 & 0.15 & 93.51 $\pm$ 23.53 & 0.26 $\pm$ 0.15 & 0.95 $\pm$ 0.21 & 0.39 & 13 & 13 \\ 
 33           & 00033 & 8.55 & 0.33 & 48.78 $\pm$ 9.98 & 0.25 $\pm$ 0.13 & 0.95 $\pm$ 0.19 & 0.41 & 15 & 15 \\ 
 34           & 00034 & 8.51 & 0.15 & 114.12 $\pm$ 43.76 & 0.04 $\pm$ 0.02 & 0.95 $\pm$ 0.24 & 0.17 & 21 & 22 \\ 
 35           & 00035 & 8.50 & 0.15 & 143.02 $\pm$ 55.51 & 0.03 $\pm$ 0.03 & 0.95 $\pm$ 0.21 & 0.34 & 11 & 11 \\ 
 36           & 00036 & 8.46 & 0.15 & 102.44 $\pm$ 31.81 & 0.05 $\pm$ 0.02 & 0.95 $\pm$ 0.19 & 0.21 & 10 & 10 \\ 
 36           & 00036 & 8.46 & 0.15 & 92.34 $\pm$ 39.98 & 0.06 $\pm$ 0.05 & 0.95 $\pm$ 0.25 & 0.30 & 6 & 6 \\ 
 38           & 00038 & 8.32 & 0.15 & 114.16 $\pm$ 28.20 & 0.05 $\pm$ 0.01 & 0.95 $\pm$ 0.16 & 0.13 & 14 & 14 \\ 
 40           & 00040 & 7.00 & 0.15 & 95.55 $\pm$ 17.94 & 0.32 $\pm$ 0.09 & 0.95 $\pm$ 0.16 & 0.26 & 12 & 12 \\ 
 40           & 00040 & 7.00 & 0.15 & 107.07 $\pm$ 19.23 & 0.29 $\pm$ 0.10 & 0.95 $\pm$ 0.16 & 0.25 & 23 & 23 \\ 
 41           & 00041 & 7.12 & 0.10 & 198.74 $\pm$ 61.71 & 0.05 $\pm$ 0.07 & 0.95 $\pm$ 0.20 & 0.13 & 19 & 19 \\ 
 45           & 00045 & 7.46 & 0.07 & 181.92 $\pm$ 59.39 & 0.05 $\pm$ 0.03 & 0.95 $\pm$ 0.21 & 0.14 & 11 & 11 \\ 
 47$^\dagger$ & 00047 & 7.84 & 0.16 & 107.18 $\pm$ 33.79 & 0.07 $\pm$ 0.03 & 0.95 $\pm$ 0.20 & 0.28 & 14 & 14 \\ 
 48$^\dagger$ & 00048 & 6.90 & 0.15 & 165.38 $\pm$ 41.80 & 0.06 $\pm$ 0.03 & 0.95 $\pm$ 0.15 & 0.42 & 11 & 11 \\ 

 \enddata
 \label{tab:mbadup}
 \end{deluxetable}

\begin{figure}[h!]
  \caption{Histogram of MBA diameters (top) and albedos (bottom) measured in this work (blue), and values for the same objects measured by the NEOWISE team previously (green). The two albedo peaks are due to the predominance of bright S-type ($p_V=0.25$) and dark C-type ($p_V=0.06$) objects in the Main Belt.}
  \centering
  \includegraphics[width=0.7\textwidth]{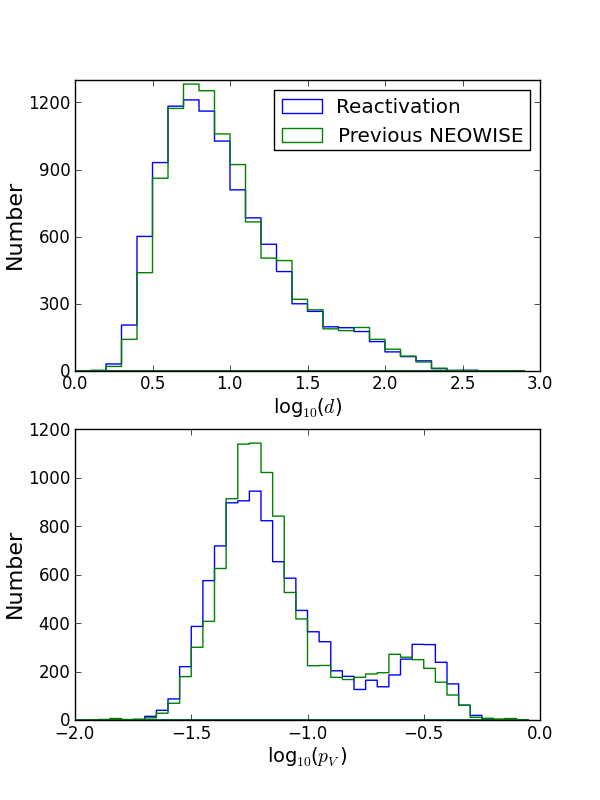}
  \label{fig:adcomp}
\end{figure}

Results were compared to previous work by the NEOWISE team \citep{Mainzer2011d,Masiero11,Mainzer12,Masiero12}. Figure \ref{fig:adcomp} shows the comparison between diameter and albedo measurements of MBAs. As observed in \citet{Masiero11}, asteroids in the Main Belt group into bright and dark types, with a greater fraction of bright objects found in the inner regions of the belt. Objects that were also modeled with the thermophysical model of \citet{Wright07} are given in Table \ref{tbl:Wright}.
\begin{deluxetable}{rcc}
\tabletypesize{\scriptsize}
\tablecaption{Measured diameters and albedos for three objects using the model of \citet{Wright07}.\label{tbl:Wright}}
\tablewidth{0pt}
\tablehead{
	\colhead{Name} & \colhead{$D$ (km)} & \colhead{$p_V$} }                                                                                      
\startdata
68267           & 0.89 $\pm$ 0.27 & 0.38 $\pm$ 0.32 \\
138127          & 0.94 $\pm$ 0.15 & 0.35 $\pm$ 0.08 \\
285944          & 1.37 $\pm$ 0.23 & 0.34 $\pm$ 0.08 \\
\enddata
\end{deluxetable}

\begin{figure}[h!]
  \caption{Top: Comparison of radar-derived diameters and albedos to the values derived in this paper. Dashed red line is shows a 1:1 relation. Bottom: $\%\Delta d$ (left) and $\%\Delta p_V$ (right) are the fractional differences between the NEOWISE Reactivation radar-derived diameters and albedos, respectively.  Dashed red line is best-fit Gaussian, with the fitted $\sigma$ given in the legends. }
  \centering
    \includegraphics[width=1.0\textwidth]{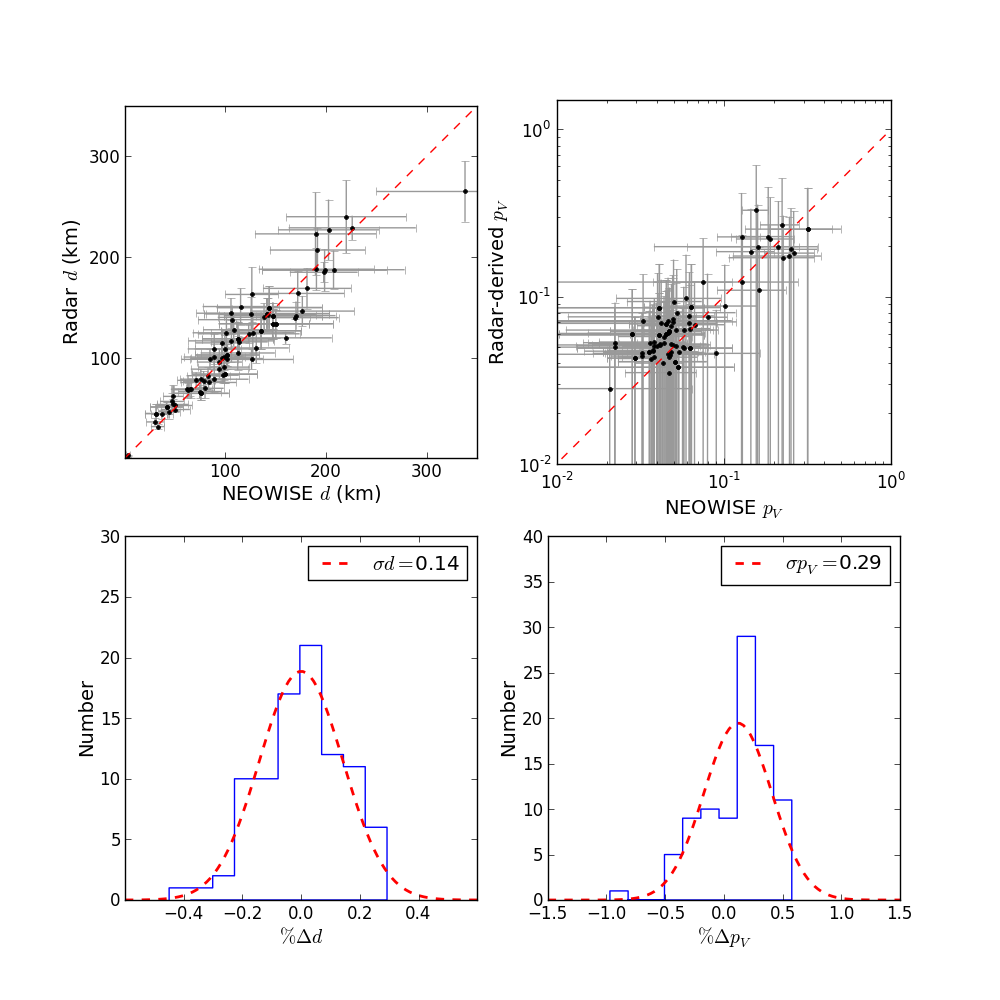}
    \label{fig:radar}
\end{figure}

When possible, derived diameters were compared to diameter measurements made from radar data. Radar-derived diameters are ideal for this purpose, as they are derived via an independent method \citep{BennerAstIV_temp}. This comparison is shown in Figure \ref{fig:radar}. Although the histograms in the figure are not perfectly Gaussian, a best-fit Gaussian to their forms gives fitted $\sigma$ values, which indicates a $14\%$ relative accuracy in diameter, and a $29\%$ relative accuracy in albedo. These values are consistent with previous NEOWISE 3-band data results \citep{Mainzer12, Masiero12}. From this comparison to radar-derived diameters and previous work, we conclude that diameters are determined to an accuracy of $\sim20\%$ or better. If good-quality H magnitudes are available, albedos can be determined to within $\sim40\%$ or better.

Roughly $3\%$ of objects in this work have significantly different derived diameters than previously published NEOWISE values. It is possible that some of these objects are elongated. NEOWISE collects a sparsely-sampled lightcurve for each object, and for example, it is possible that the prime mission happened to observe these objects in a more edge-on shape, whereas the reactivation observations tended to observe a wider side of the object. 
Alternatively, changes in viewing geometry between epochs could result in different diameter measurements; a pole-on viewing geometry could have a larger cross section than a geometry aligned with the plane of the equator.

For a small ($<1\%$) fraction of objects, diameter fits could not reproduce optical magnitudes for a realistic range of albedos. This may be due to a large light curve amplitude (see column W2 amp. for the amplitude of the 3.4 $\mu$m band light curve, though note that this is a sparsely sampled light curve), uncertainty in $G$ slope values used to derive $H$ magnitudes, or other reasons noted in \citet{Mainzer2011d,Masiero11,Mainzer12,Masiero12}. Poor-quality $H$ values can drive albedo fits to extremes; therefore  very low ($\sim 0.01$) measurements may be signs of this phenomenon.

\begin{figure}[h!]
  \caption{NEOWISE detects large NEAs, and discoveries tend to be dark. Cyan circles are measured diameters and albedos of objects detected in the first year of NEOWISE's Reactivation mission; black squares indicate NEAs discovered by NEOWISE. Error bars on detected objects omitted for clarity.}
  \centering
  \includegraphics[width=0.7\textwidth]{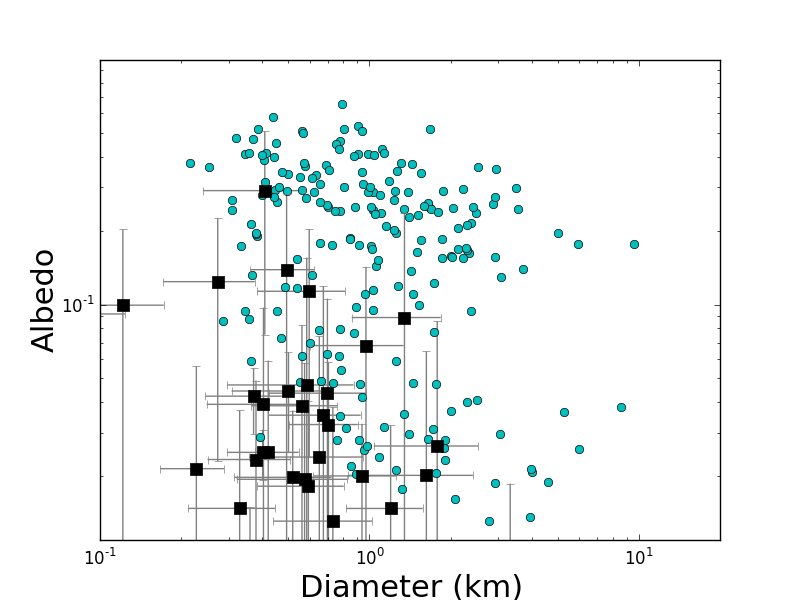}
  \label{fig:disc}
\end{figure}

We have plotted the diameters and albedos of NEOWISE Year One Reactivation discoveries, along with all NEAs detected by NEOWISE (Figure \ref{fig:disc}). The trend observed in \citet{14MainzerRestart} is also present here: NEOWISE tends to discover darker NEAs than optical surveys. This is a direct consequence of the infrared wavelengths that NEOWISE employs.

\subsection{NHATS Targets}

Five objects in this paper meet the  NEO Human Space Flight Accessible Targets Study  qualifications \citep{BarbeeNHATS}. These objects are listed in Table \ref{tbl:NHATS}. If an object was observed over multiple epochs, values of $d$ and $p_V$ in this table are the averages of the values and associated errors derived at each of those epochs. Asteroid 419624 was discovered in 2010 by NEOWISE.

\begin{deluxetable}{rccc}
\tabletypesize{\scriptsize}
\tablecaption{Measured diameters and albedos for objects that meet NHATS criteria. Also included are the minimum round trip time in days, as determined by the \citet{BarbeeNHATS} study.\label{tbl:NHATS}}
\tablewidth{0pt}
\tablehead{
	\colhead{Name} & \colhead{$D$ (km)} & \colhead{$p_V$} &  \colhead{Minimum round trip (days)} }                                                                                       
\startdata

1943 Anteros & 2.30 $\pm$ 0.05 & 0.17 $\pm$ 0.02 & 354 \\ 
35107        & 1.00 $\pm$ 0.15 & 0.34 $\pm$ 0.10 & 354 \\
363505       & 1.90 $\pm$ 0.05 & 0.03 $\pm$ 0.01 & 314 \\
387733       & 0.33 $\pm$ 0.05 & 0.44 $\pm$ 0.15 & 354 \\
419624       & 0.35 $\pm$ 0.13 & 0.09 $\pm$ 0.15 & 362 \\
\enddata
\end{deluxetable}

\section{Conclusion}

We present preliminary diameters and albedos for 7,959 asteroids observed in the first year of the NEOWISE Reactivation mission. Five of these objects are NHATS targets. Future work by the NEOWISE team includes preliminary characterization results from the continuing mission. 

Uncertainties on $d$ and $p_V$ are consistent with the errors measured during the initial post-cryo mission. NEOWISE is expected to maintain this pace of detection and NEO discovery for the extent of its mission, currently expected to run through 2017. These results demonstrate the power of infrared survey telescopes to characterize basic physical parameters for large numbers of small bodies.

\section{Acknowledgments}

CRN was partially supported by an appointment to the NASA Postdoctoral Program at the Jet Propulsion Laboratory (JPL), administered by Oak Ridge Associated Universities through a contract with NASA. This publication makes use of data products from the Wide-field Infrared Survey Explorer, which is a joint project of the University of California, Los Angeles, and JPL/California Institute of Technology, funded by NASA. This publication also makes use of data products from NEOWISE, which is a project of the JPL/California Institute of Technology, funded by the Planetary Science Division of NASA. This research has made use of the NASA/IPAC Infrared Science Archive. The JPL High-Performance Computing Facility used for our simulations is supported by the JPL Office of the CIO.

 This project used data obtained with the Dark Energy Camera (DECam), which was constructed by the Dark Energy Survey (DES) collaboration.
Funding for the DES Projects has been provided by 
the U.S. Department of Energy, 
the U.S. National Science Foundation, 
the Ministry of Science and Education of Spain, 
the Science and Technology Facilities Council of the United Kingdom, 
the Higher Education Funding Council for England, 
the National Center for Supercomputing Applications at the University of Illinois at Urbana-Champaign, 
the Kavli Institute of Cosmological Physics at the University of Chicago, 
the Center for Cosmology and Astro-Particle Physics at the Ohio State University, 
the Mitchell Institute for Fundamental Physics and Astronomy at Texas A\&M University, 
Financiadora de Estudos e Projetos, Funda{\c c}{\~a}o Carlos Chagas Filho de Amparo {\`a} Pesquisa do Estado do Rio de Janeiro, 
Conselho Nacional de Desenvolvimento Cient{\'i}fico e Tecnol{\'o}gico and the Minist{\'e}rio da Ci{\^e}ncia, Tecnologia e Inovac{\~a}o, 
the Deutsche Forschungsgemeinschaft, 
and the Collaborating Institutions in the Dark Energy Survey. 
The Collaborating Institutions are 
Argonne National Laboratory, 
the University of California at Santa Cruz, 
the University of Cambridge, 
Centro de Investigaciones En{\'e}rgeticas, Medioambientales y Tecnol{\'o}gicas-Madrid, 
the University of Chicago, 
University College London, 
the DES-Brazil Consortium, 
the University of Edinburgh, 
the Eidgen{\"o}ssische Technische Hoch\-schule (ETH) Z{\"u}rich, 
Fermi National Accelerator Laboratory, 
the University of Illinois at Urbana-Champaign, 
the Institut de Ci{\`e}ncies de l'Espai (IEEC/CSIC), 
the Institut de F{\'i}sica d'Altes Energies, 
Lawrence Berkeley National Laboratory, 
the Ludwig-Maximilians Universit{\"a}t M{\"u}nchen and the associated Excellence Cluster Universe, 
the University of Michigan, 
{the} National Optical Astronomy Observatory, 
the University of Nottingham, 
the Ohio State University, 
the University of Pennsylvania, 
the University of Portsmouth, 
SLAC National Accelerator Laboratory, 
Stanford University, 
the University of Sussex, 
and Texas A\&M University.

This work makes use of observations from the LCOGT network.

Follow-up included observations obtained at the Gemini Observatory, which is operated by the Association of 
Universities for Research in Astronomy, Inc., under a cooperative agreement with 
the NSF on behalf of the Gemini partnership: the National Science Foundation 
(United States), the National Research Council (Canada), CONICYT (Chile), the 
Australian Research Council (Australia), MinistŽrio da Cincia, Tecnologia e 
Inova‹o (Brazil) and Ministerio de Ciencia, Tecnolog'a e Innovaci—n Productiva 
(Argentina).

We thank the anonymous referee for their thoughtful and thorough consideration
of our manuscript.

\clearpage

\notetoeditor{Complete Tables \ref{tab:obsshort}, \ref{tab:mbanew}, and \ref{tab:mbadup} should appear in online supplementary material.}

\end{document}